\documentclass[a4paper,fleqn]{cas-dc}

\usepackage[authoryear]{natbib}
\usepackage{graphicx}
\usepackage{amsmath}
\usepackage{array}
\usepackage{caption}  
\usepackage{booktabs}
\usepackage{tabularx}
\usepackage{multirow}
\usepackage{hyperref}
\usepackage{placeins}
\usepackage{orcidlink}
\usepackage{float}
\usepackage{gensymb}
\usepackage{enumitem}
\usepackage{hyperref}
\usepackage{breakurl}
\usepackage[intoc]{nomencl}
\usepackage{tcolorbox}
\makenomenclature

\begin{document}
\let\WriteBookmarks\relax
\def\floatpagepagefraction{1}
\def\textpagefraction{.001}

\nomenclature{$A_i$}{Area under the $AI(t)$ curve for entity $i$ [$A_i \in \{0, \dots, \infty\}$]}
\nomenclature{$i$}{Geographic entity index [$i \in \{1, \dots, N\}$]}
\nomenclature{$A_m$}{Area under the $AI(t)$ curve of the regional mean [$A_m \in \{0, \dots, \infty\}$]}
\nomenclature{$\tilde{A}_i$}{Normalized $A_i$ [$\tilde{A}_i \in \{0, \dots, \infty\}$]}
\nomenclature{$\tilde{A}_m$}{Normalized $A_m$ [$\tilde{A}_m = 100$]}
\nomenclature{$d$}{Degree of the polynomial function [$d \in \{2, \dots, T{-}1\}$]}
\nomenclature{$t$}{Time point (years) [$t \in \{t_{k=1}, \dots, t_{k=T}\}$]}
\nomenclature{$k$}{Time points index in the time series [$k \in \{1, \dots, T\}$]}
\nomenclature{$T$}{Number of time points in the time series [$T \in \{1, \dots, \infty\}$]}
\nomenclature{$t_{k=1}$}{Initial time point of the timeline (years)}
\nomenclature{$t_{k=T}$}{Final time point of the timeline (years)}
\nomenclature{$C_i$}{Best-fitting $AI(t)$ curve for entity $i$ [$C \in \text{curves set}$]}
\nomenclature{$C_m$}{Best-fitting $AI(t)$ curve of the regional mean [$C \in \text{curves set}$]}
\nomenclature{$j$}{Intersection point index between $C_i$ and $C_m$ [$j \in \{1, \dots, n\}$]}
\nomenclature{$t_{i,j}$}{Time of intersection point $j$ between $C_i$ and $C_m$ for entity $i$ (years) [$t_{i,j} \in \{t_{i,j=1}, \dots, t_{i,n}\}$]}
\nomenclature{$t_{i,j+1}$}{Time of the next intersection point after $t_{i,j}$ (years) [$t_{i,j+1} \in \{t_{j=2}, \dots, t_{n+1}\}$]}
\nomenclature{$\alpha_{i,j}$}{Angle between $C_i$ and $C_m$ for entity $i$, measured after intersection point $j$ [$\alpha_{j=n} \in \{0^\circ, \dots, 360^\circ\}$]}
\nomenclature{$S_{i,j}$}{Feedback factor for entity $i$ at intersection point $j$ [$S_{i,j} \in \{0, 1\}$]}
\nomenclature{$\Delta_i$}{Difference between $\tilde{A}_i$ and $\tilde{A}_m$ [$\Delta_i \in \{-100, \dots, \infty\}$]}
\nomenclature{$F_{i,t_{k=1}}$}{Feedback duration for entity $i$ from $t_{k=1}$ to $t_{j=1}$ (years) [$F_{t_{k=1}} \in \{0, t_{j=1}-t_{k=1}\}$]}
\nomenclature{$F_{i,j}$}{Feedback duration for entity $i$ between intersection points $j$ and $j+1$ (years) [$F_{i,j} \in \{0, t_{j+1}-t_{j}\}$]}
\nomenclature{$PVa_{i,t}$}{Total installed PV area in entity $i$ at time $t$ ($m^2$) 
[$PVa_{i,t} \in \{0, \dots, Ba_i\}$]}
\nomenclature{$Ba_i$}{Built-up area of entity $i$ ($m^2$) [$Ba_i \in \{0, \dots, \infty\}$]}
\nomenclature{$10^6$}{Scaling constant}
\nomenclature{$AI_{i,t}$}{Adoption intensity for entity $i$ at time $t$ [$AI_{i,t} \in \{0, \dots, 10^6\}$]}
\nomenclature{$ATI_i$}{Adoption over time index for entity $i$ [$ATI_i \in \{0, \dots, \infty\}$]}
\nomenclature{$AI_{m,t}$}{Mean $AI$ across all entities at time $t$ [$AI_{m,t} \in \{0, \dots, 10^6\}$]}
\nomenclature{$LAI_i$}{Normalized $AI_i$ at time point $t_T$ [$LAI_i \in \{0, \dots, \infty\}$]}
\nomenclature{\textit{Entry-time\textsubscript{i}}}{Time of entity $i$ crossing a defined threshold share of $AI_{m,t}$ (years) [$\in \{0, \dots, T\}$]}
\nomenclature{\textit{Latest-trajectory\textsubscript{i}}}{Angle following the last intersection point ($\alpha_{j=n}$) of entity $i$ [$\alpha_{j=n} \in \{0^\circ, \dots, 360^\circ\}$]}

\shorttitle{A Typology of Renewable Energy Adoption Trajectories over time}

\shortauthors{Blushtein-Livnon et~al.}

\title [mode = title]{Beyond Leaders and Laggards: A Typology of Renewable Energy Adoption Trajectories with Evidence from Off-Grid Communities}                      

\author[1]{Roni~Blushtein-Livnon}[orcid=0000-0002-3493-4894]
\fnmark[1]
\fntext[fn1]{Corresponding author.
E-mail address: livnon@bgu.ac.il}

\author[1,2]{Tal~Svoray}[orcid=0000-0003-2243-8532]
\author[3]{Itay~Fischhendler}[orcid=0000-0001-5355-9316]
\author[4]{Havatzelet~Yahel}[orcid=0000-0002-9956-0491]
\author[4]{Emir~Galilee}[orcid=0000-0001-7892-4172] 
\author[1]{Michael~Dorman}[orcid=0000-0001-6450-8047]

\affiliation[1]{organization={Department of Environmental, Geoinformatics and Urban Planning Sciences, Ben-Gurion University of the Negev}, country={Israel}}
\affiliation[2]{organization={Department of Psychology, Ben-Gurion University of the Negev}, country={Israel}}
\affiliation[3]{organization={Department of Geography, The Hebrew University of Jerusalem}, country={Israel}}
\affiliation[4]{organization={Ben Gurion Institute for the Study of Israel and Zionism, Ben-Gurion Israel Research Institute},
country={Israel}}

\begin{abstract}
Understanding the dynamics of renewable energy adoption is critical for designing strategies that accelerate its deployment. This is an urgent priority for achieving global climate targets and improving human well-being, particularly in marginalized off-grid regions affected by energy poverty. However, existing research on the diffusion of new technologies lacks systematic approaches for distinguishing between adoption trajectories over time. This paper presents a time-series-based analytical framework, applicable at any spatial context and scale, for quantifying the adoption dynamics of geographic entities within a given region and classifying each according to its behavior over time. The framework introduces a novel index, ATI, designed to measure cumulative adoption intensity and capture shifts in adoption trends. By doing so, ATI improves the ability to distinguish between fundamental adoption trajectories over time. The framework integrates ATI with three pivotal adoption features, shown to be indicative of adoption dynamics, and constructs a comprehensive typology of adoption paths. This typology consists of eight distinct adoption paths, including two previously unrecognized trajectories: the \textit{decelerating} path and the \textit{declining moderate} path. Applying the framework to a case study of an off-grid Bedouin population in southern Israel revealed substantial proportions of these two retreating trends. Identifying these two paths is crucial for preventing backsliding and addressing stagnation in the diffusion process. The leaping path, however, was found to be negligible. Additionally, behavioral differences exist within both front-runner and trailing groups. Distinguishing between them may improve the effectiveness of acceleration measures by aligning them with each trajectory’s specific characteristics and needs.
\end{abstract}



\maketitle

\begin{figure*}[!t]
\renewcommand{\nomname}{Nomenclature}
\renewcommand{\nompreamble}{\vspace{0.5em}}

\footnotesize
\setlength{\nomitemsep}{2.8pt}
\setlength{\nomlabelwidth}{2.2cm}
\renewcommand{\nomlabel}[1]{\hspace{0.4em}#1\hfill}

\makebox[\textwidth][c]{%
  \begin{minipage}{0.88\textwidth}
    \begin{tcolorbox}[
        colback=gray!3,
        colframe=black,
        boxrule=0.5pt,
        arc=0pt,
        left=6pt, right=6pt, top=6pt, bottom=6pt
    ]
    \printnomenclature
    \end{tcolorbox}
  \end{minipage}
}
\vspace{-1em}
\end{figure*}

\section{Introduction}
Renewable energy (RE), particularly solar energy, has become a cost-competitive power source, driven by large-scale manufacturing and a sharp decline in PV prices. As of 2024, RE supplies over 7\% of the global electricity \citep{IRENA2020Stats, iea2025electricity}. Nevertheless, significant spatial differences persist in RE adoption, varying in intensity, initiation timing, progression rates, and duration for full transition \citep{grubler2016apples, sovacool2016long}. This heterogeneity is attributed to various factors, including climatic conditions \citep{lamp2023sunspots}, regulatory and policy frameworks \citep{best2018adoption}, socio-economic characteristics \citep{balta2021energy}, urban attributes \citep{kosugi2019neighborhood, dharshing2017household}, spatial spillover, or peer effect \citep{graziano2015spatial}, exposure to information and social interactions \citep{carattini2018social}, and interpersonal characteristics \citep{wolske2017explaining}, which are particularly relevant given that adoption decisions ultimately occur at the household level.

Disparities in both actual RE adoption rates and the underlying propensity to adopt, were observed across multiple spatial contexts and scales, ranging from differences between the global- north and south \citep{TrackingSDG72024} to variations across countries \citep{morcillo2022assessing, baldwin2019countries}, administrative counties and in regional domains \citep{muller2020spatial, macintyre2021spatial}, municipalities \citep{bokanyi2022urban, rigo2024explain}, and at the neighborhood level \citep{kosugi2019neighborhood, abbar2018modeling}. 

These disparities often highlight fundamental inequalities in electricity access, especially in off-grid and marginalized areas, where reliable and affordable power remains out of reach, even as solar panel adoption continues to expand significantly. As of 2024, an estimated 685 million people still lack access to electricity \citep{TrackingSDG72024}. These inequities, in turn, perpetuate broader socioeconomic disadvantages and have become a central focus of energy justice research \citep{lacey2020energy}, with a growing recognition of the pivotal role electricity access plays in poverty reduction and household well-being \citep{batinge2019perpetuating, saha2025empowering}. Studies of off-grid societies provide a compelling illustration of the transformative impact that electricity access can have. This includes enhancement of household consumption; income generation and employment opportunities; health and education outcomes; overall socioeconomic progress \citep{bridge2016electricity, KHANDKER20121}; and, finally, women's empowerment through the reduction of heavy labor burdens \citep{PUEYO2019170}. Given the role of electricity access in socioeconomic development, and the necessity of RE in bridging the gap between current emission trajectories and the global target of limiting warming to below 2°C, accelerating the deployment of RE technologies, particularly PV systems, is of critical importance \citep{UNEP2022}.

Achieving rapid and equitable adoption of RE technologies requires a deeper understanding, beyond existing knowledge of how adoption processes unfold across diverse contexts. Systematic temporal measurement of adoption processes allows for the identification of evolving statuses and dynamics of different spatial entities, along the adoption trajectory, and to assess both their adoption potential and the means by which it can be further accelerated. Such analyses are essential not only for mainstream regions with existing infrastructure, but also, perhaps even more so, for off-grid regions, where adoption trajectories are often shaped by distinct constraints, including limited access to finance \citep{ kizilcec2021examining, barry2020pay}, weak institutional support \citep{hyun2021modeling}, and socio-economic exclusion \citep{simpson2021adoption}. In these contexts, PV deployment serves not only as an energy source but as a catalyst for welfare gains, productive uses, and localized resilience \citep{diallo2020effects, aarakit2022role}.

Analyzing adoption temporal dynamics, within their regional context, can thus reveal both general diffusion patterns and context-specific barriers or enablers, such as key local hubs that may stimulate imitation and reinforcement, or marginalized groups that face limited access to energy. These insights are critical for designing effective and targeted policy interventions to promote widespread adoption and maximize deployment, particularly in areas where electrification is both an urgent challenge and a transformative opportunity.

The current study introduces an analytical framework for systematically capturing the temporal dynamics of RE adoption across geographical entities and classifying them by their trajectories, relative to regional trends. The following sections explore the prevailing methods and measures used to assess RE adoption and the typologies applied to categorize adoption patterns and adopter groups. Following this, the study’s contributions to the literature are outlined, emphasizing how the proposed framework enhances the understanding of RE adoption dynamics.

\subsection{Quantifying Adoption}

An adoption process of new technology over time is commonly depicted as an S-shaped curve, illustrating its cumulative trajectory from initiation to saturation, and capturing the non-linear nature of diffusion \citep{young1993technological, schilling2009technology}. Various adaptations of this curve, such as the Bass model \citep{guidolin2010cross, ratcliff2016using}, Richards growth model \citep{marinakis2012forecasting}, and Gompertz curve \citep{gutierrez2005forecasting}, were developed to account for context-specific factors that shape adoption trajectories. These models differ in several key characteristics, including their flexibility, growth rate, symmetry, and the timing of their tipping and inflection points. Despite these differences, all models share a common perspective: Adoption is quantified as a cumulative process that continues until saturation is reached. Moreover, all models incorporate a catalytic tipping point that marks the onset of rapid growth, and an inflection point, beyond which the growth rate begins to decline until adoption approaches full saturation \citep{bento2016measuring}.

Previous research focused on identifying key milestones along these adoption curves to characterize and compare adoption processes across entities. These milestones typically include timing of initiation \citep{gort1982time, milliou2011timing, overby2019adopters}, end of formation phase \citep{bento2016measuring, wilson2011lessons}, timing of tipping point \citep{lenton2022operationalising}, and the duration required to reach adoption saturation \citep{sovacool2016long, grubler2016apples, andersen2023faster}. Using these milestones, several studies aimed to explain disparities in adoption intensity between entities, primarily at the national level \citep{bate2023determinants, baldwin2019countries, brutschin2021failing}. 

Entry times of entities into an adoption process were commonly identified as significant milestones to distinguish between developed/central regions and developing/peripheral regions (e.g., \citep{qiu2020evolution, bunea2020adoption, duan2018peer}). The latter typically experience delayed entries. These differences may stem from various factors, including the context at the entry point, such as potential profitability of technology, or accumulated technological knowledge \citep{aghion2015knowledge, fong2009technology}. 

For instance, as knowledge accumulates over time, late adopters are able to capitalize on gained expertise. This provides them with access to information and insights that were unavailable during initial adoption phases, thereby potentially accelerating their adoption process \citep{qiu2020evolution}. These findings point to the possibility that high levels of adoption can be attained through various pathways, such as leapfrogging, which enables late adopters to bypass intermediate stages of the process \citep{lee2020economics, duan2018peer}. Multiple studies confirmed this, showing that adoption duration can be shorter for developing regions, compared with developed regions that have started to adopt earlier \citep{wilson2009meta, wilson2011lessons}. However, the ability of late adopters to reach full transition remains uncertain, with ongoing debate surrounding the trajectories they are able to follow \citep{wilson2011lessons, wilson2012up}. \citet{gosens2017faster}, for example, has found that initiation time can act as both, an enabler and a barrier, where late adopters may benefit from faster learning and can complete the adoption process more rapidly, whereas latest adopters may not learn as efficiently, which can lead to delays in the overall process and potentially result in lower adoption rates.

Another uncertainty revolves around the tipping point and the extent to which it guarantees widespread adoption \citep{azmat2023convergence, lenton2022operationalising}. Tipping point is generally viewed as an irreversible shift, where self-reinforcing positive feedback loops propel the diffusion process forward, forcing a change once the tipping point is crossed \citep{Milkoreit_2018}. This suggests that after the tipping point, adoption accelerates steadily as more adopters deploy the technology. This approach was challenged, and studies highlight how adoption trajectories might be changed throughout the process \citep{lockwood2022, edmondson2019}. Recently, \citet{geels2023socio} have argued that post-tipping point dynamics may include not only positive, but also negative feedback, that can decelerate, halt, or even reduce spread by diffusion. For example, in a PV deployment process in the UK and Germany, after the tipping point was reached, policy shifts and public debates over costs led to negative feedback that slowed adoption rates and changed the trend \citep{ayoub2024}. This suggests that diffusion is not necessarily a one-way trajectory, as broader contexts can activate reactions that counteract previously self-reinforcing growth. Therefore, quantifying an ongoing, continuous adoption process bears important information compared with an approach relying solely on milestone-based assessments, such as tipping point timing. 

Saturation phase timing is also an inconsistent metric for cross-entity comparison. While saturation is often used as a marker of maturity in adoption processes, its measurement remains challenging across different entities and contexts. For example, even when an entity reaches maturity more rapidly by leapfrogging earlier stages, its adoption intensity may still differ significantly from that of other entities \citep{griliches1957hybrid, grubler1990rise}. Comin and Mestieri showed that although the time lag between early and late adopters of new technologies tends to decrease, thanks to faster learning and globalization, the actual extent to which different groups adopt and use these technologies may become increasingly unequal \citep{comin2018if}. That is, late adopters may not catch up with early adopters and continue to utilize the technology at lower levels. This is critical in the context of adopting RE for climate mitigation, as rapid saturation does not ensure that the intensity at this phase is sufficient to achieve the targeted climatic goals \citep{mi2024research}. Moreover, this metric is inherently limited, as measuring adoption duration from initiation to saturation is only feasible retrospectively, after the diffusion process has run its full course.

Still, despite extensive research on RE adoption, the complexity and variation of adoption processes remain insufficiently understood. Unlike studies examining technologies whose diffusion has already reached completion (e.g., \citep{van2015comparing, grubler1997time}, analyzing an ongoing adoption process presents significant challenges, particularly in identifying key milestones along its trajectory. There are some studies that investigate actual adoption rates of real-time processes by using time series data (e.g., \citep{Morcillo2022, cherp2021national, zielonka2023probabilistic}). The majority of them explore different approaches, such as using entry time as a comparative metric \citep{milliou2011timing, pfeiffer2013explaining, subtil2014international}, assessing conclusions of the technology formative phase \citep{bento2016measuring, brutschin2021failing}, or conducting a cross-sectional comparison of adoption levels at selected time points \citep{autant2010measuring}. However, analyzing adoption behavior by relying on a limited set of milestones, rather than comprehensively tracking the process's continuous course, reduces the analytical resolution needed to distinguish between fundamental adoption dynamics. These research gaps highlight the need for a robust analytical framework that integrates granular time series data and captures all relevant factors and variations throughout the process. Such an approach enables more precise quantification and a deeper understanding of the underlying dynamics, allowing for clear differentiation between distinct adoption paths.

\subsection{Typology of adoption types}
There is broad recognition of the importance of adopter typologies as a tool for analyzing adoption processes, and as a foundation for developing strategies to promote it \citep{wang2022, hoicka2025insights}. In a significant portion of the literature, classifying adopter types relies on Everett M. Rogers' seminal work on the diffusion of innovations \citep{deugerli2015analyzing, wang2022, franceschinis2017adoption}. While some alternative categorization approaches exist, most commonly used typologies tend to classify geographic entities, or markets, based on indicators related to adoption propensity, innovativeness, or policies aimed at accelerating RE adoption \citep{knill2012really, döme_2023} rather than monitoring actual adoption behavior, namely de facto deployment or usage. Others focus on an individual level by measuring attitudes, awareness, or willingness to adopt \citep{wang2022, schulte2022meta, islam2014household}.

When classifying geographic entities, studies often employ generalized distinctions, primarily differentiating between leaders or laggards, sometimes including intermediate groups such as followers or pushers \citep{tarpani2022energy, andresen2002leaders, tobin2017leaders}. Another common classification is based on the Bass diffusion model \citep{guidolin2010cross, Lan02012020}, which differentiates between innovators and imitators by identifying inflection points along the process \citep{mahajan1990determination}, namely time points when adoption rates shift substantially. However, such classifications can mostly be applied retrospectively to processes that reached maturity.

While identifying fundamental adopter types is essential for promoting future adoption, especially since adoption processes are strongly influenced by interactions between these groups, such classifications are often insufficient. They fail to capture finer intermediate adopter types that may play a crucial role in shaping the region's adoption trajectory. For example, imitation-based adoption, widely recognized as a key mechanism in innovation diffusion \citep{massiani2015choice, ratcliff2016using}, does not necessarily stem from mimicking leaders or innovators, who are often driven by ideology or an inherent affinity for technological innovation rather than economic considerations, and may not serve as a relatable influence for potential adopters. This suggests that distinguishing between ultimate leaders and those who enter the process slightly later, once its economic viability is established, is important since the latter can provide a more compelling source of imitation for the conservative majority \citep{RUOKAMO2023103183}. Similarly, identifying leapfrogging adopters could also be essential for understanding regional dynamics, as they may serve as a success case that may inspire laggards. Despite Rogers' highly detailed classification, his framework does not account for these types. Furthermore, it does not allow the identification of adopters whose process slows down, stalls, or even declines throughout the timeline, overlooking an important dimension of real-world adoption patterns \citep{geels2023socio, ayoub2024}.

Identification of the existing diversity of adopter groups is essential for promoting transitions, especially since adoption processes are influenced by between-group dynamics \citep{bass1969new, baptista1999diffusion, goldenberg2000marketing}. Identifying Influencers may allow for leveraging their impact to promote adoption and act as a catalyst to reduce spatial disparities. Moreover, a typology of various pathways and changes, unfolding throughout their progression, would facilitate a deeper understanding of their effect on adoption intensities across time and space. The ability to accurately quantify adoption processes, including shifts in temporal trajectories occurring within it over time, even after tipping or inflation points, may provide a solid foundation to distinguish between various paths along the process and define a typology that enables a discerning distinction between fundamental adopter groups.

As was aforementioned, despite extensive RE adoption research, significant gaps remain in the understanding of its complex dynamics. First, mostly due to low data resolution at the regional/local scale, existing analyses often rely on a limited set of discrete features, failing to capture the continuous and evolving nature of adoption trajectories over time. Furthermore, how these characteristics influence RE adoption processes in their entirety, including the role of initial adoption timing in shaping subsequent diffusion patterns, and whether adoption pathways are inherently unidirectional, remains poorly understood. Additionally, current adopter typologies are overly coarse and lack a foundation in the process dynamics that may shape adoption behavior. Consequently, they fail to account for critical variations such as deceleration, stagnation, or even decline in adoption levels. Addressing these gaps requires an analytical framework that systematically quantifies adoption dynamics, identifies crucial features along the trajectory, and provides a comprehensive classification of adoption pathways.

\subsection{Research Contribution}
To address the research gaps, this study makes the following key contributions:
\begin{enumerate} [leftmargin=1.5em, itemsep=0pt]

\item We developed an analytical framework for systematic examination and interpretation of adoption processes, achieved through extracting a comprehensive typology of adoption dynamics, for geographic entities, within a given region. Grounded in theoretical foundations, our framework brings together four key measurable adoption features, a structured profiling procedure, and a rule-based classification. Integration of these components, each derived from rigorous quantification of the unfolding process, establishes a robust basis for identifying diverse adoption paths. A core pillar of this framework is a newly developed index: the Adoption over Time Index (ATI), which allows capturing both process intensity and its dynamic nature. 

\item Identification of new adoption paths for a more profound understanding of adoption dynamics. Our new typology has enabled to define two previously unclassified adopter types: the \textit{decelerating path} and the \textit{declining moderate path}. Both show distinct patterns of either stagnation or decline in their adoption trajectories, though they differ in intensity. A pinpoint of deteriorating entities, that could not be defined otherwise, is essential for targeted interventions to enhance RE technology penetration. Our typology also defines a third adopter type, \textit{leaping path}, which has garnered large interest in the literature but remains absent from existing typologies. That is primarily because former typologies did not account for adoption dynamics. These three paths, combined with a more fine-grained classification of well-established adoption types, form a robust and comprehensive typology that enables a refined detection of various adoption pathways and their distinct dynamic nature.

\item The suggested framework was applied to a vast region, populated by off-grid communities, using a decade-long time series of household-level small-scale PV adoption data—marking. This is, to our knowledge, the first effort to quantify adoption paths focused on off-grid communities. The framework application allows to study how adoption processes have evolved over time, delivering vital insights into the region's overall adoption profile. It facilitates identification of key imitation hubs, namely, leader areas that may enhance adoption through social networks and peer effect, and potential pockets where adoption could be accelerated through targeted interventions. It may also be used to identify distinct transitional shifts in adoption behaviors throughout the observed period. Such insights may establish a foundation for policymakers to design customized strategies tailored to the region’s unique characteristics, supporting increased PV deployment across off-grid communities.
\end{enumerate}

\section{Methodology}
This section presents an analytical framework for defining a typology of distinct adoption dynamics over time, namely \textit{adoption paths}, within a defined region comprising several geographical entities. It offers applicability across spatial scales and consists of the following three stages. \textit{First}, four selected key features are extracted to quantify adoption dynamics over time. Those are: (i) initiation timing of the process; (ii) its intensity over time; (iii) the dynamics that unfold throughout it; and (iv) its status at the most recent point in time. The integration of these features captures the complexity and nuances of the adoption process as a whole. \textit{Secondly}, for each geographical entity studied, an adoption profile is generated, based on its unique composition of the aforementioned feature values; and \textit{thirdly}, each entity’s adoption dynamic is delineated through a typology that distills and characterizes distinct adoption pathways.

\subsection{Key adoption features}
We identified the aforementioned four key features as the most essential markers across a new technology adoption process, based on established indicators widely discussed in the literature on technology adoption in general, and specifically on RE adoption, as was detailed in the literature review above (e.g., \cite{gort1982time, grubler2016apples, sovacool2016long, lenton2022operationalising, andersen2023faster, grubler1999dynamics}). A combination of these features enables a comprehensive and theoretically grounded definition of the sought-after adoption pathways. The first feature is an index developed here by us, denoted \textit{adoption over time index (ATI)}. ATI is computed using time series analyses of adoption intensity, and it accounts for dynamic trend shifts that unfold throughout adoption processes. The second feature \textit{Entry time} is the point in time at which the studied entity initiated its adoption process. The third feature \textit{Latest Adoption Intensity (LAI)} quantifies the entity's most recent adoption level within an examined timeline, providing a measure of its latest adoption state. The fourth feature,\textit{Latest trajectory}, captures the entity's last trend relative to the regional mean trend, and serves as a proxy for assessing its future progression. The following section details concepts and calculations of these features.

\subsubsection{Adoption over Time Index (ATI)} \label{sec: ATI}
ATI was developed to estimate both intensity and dynamics of the adoption process over time, for each geographic entity. Quantification of adoption intensities over time, accounting for shifts in temporal trajectories, captures differences between behavior paths. Table \ref{algorithm} is a pseudo-code for ATI computing. ATI incorporates three conceptual components: (1) area under the cumulative adoption intensity curve of each entity; (2) area under the cumulative regional intensities mean curve, serving as a benchmark for assessing the relative adoption position of each geographic entity; and (3) a feedback mechanism that refines the quantification of each entity's adoption intensity by incorporating shifts in its temporal trend. In more detail: 

\smallskip
(1) \textit{AUC of cumulative adoption} ($A_i$): Area Under the Curve (AUC), computed using integral calculus, is a well-established metric for quantifying dynamic processes. It is widely used across disciplines, including economics and psychology (e.g., to measure behavioral features or assess growth curves \citep{westbrook2013subjective, yao2024quantifying}), medical sciences and epidemiology (e.g., to track various health indicators progression \citep{pruessner2003two, dowd2009socio}), pharmacokinetics (e.g., to evaluate drug exposure \citep{dudhani2010f, jackson2022simulation}), ecology (e.g., to describe ecological dynamics and to compare different ecological disturbances effects \citep{jentsch2019theory, sun2023evaluation}). It has also been applied in a variety of other domains, such as physics \citep{barkai2014area} and statistical and machine learning \citep{carrington2022deep}. Estimating technology adoption over time using AUC is appropriate for two reasons: First, the theory of new technology diffusion represents change over time in terms of area (e.g., probabilities on a cumulative normal curve) \citep{rogers2003diffusion}. Second, since adoption intensities and time are both continuous variables, integral and differential calculus are most suitable. Accordingly, in the context of RE diffusion, AUC can serve as a robust and effective measure of cumulative adoption over time for each geographic entity (i), since it captures its time-integrated adoption intensity, which is expressed as a single and comparable value across entities. Moreover, note that since we chose to use an intensity metric defined by the total area of PV (see Section \ref{sec: AI} and Eq.~\ref{AI}), the resulting AUC can also serve as a proxy for the aggregate installed capacity.

As the foundation for cumulative curve fitting and AUC calculation, a time series metric representing the adoption intensities must be selected. We used here PV density per entity, \textit{Adoption intensity (AI)}, as a base metric (details in Section \ref{sec: AI}). The term intensity was chosen to represent adoption level, as intensity commonly denotes an amount per unit of area, time, or volume, thus, capturing not only the magnitude, but also its spatial/temporal distribution. Next, Curve fitting is performed by selecting the best fit from a set of commonly used cumulative curves in innovation adoption studies. This includes the Logistic, Gompertz, Bass, Generalized Richards, Cumulative Normal, Exponential, and Bertalanffy functions, all commonly used to quantify the adoption of new technologies over time; polynomial functions of degree, $d$, where $d \in \{2, \ldots, T-1\}$ to provide flexibility and allow for potential declines in adoption trajectory (where T is the number of time points); and a Linear function to allow for simple and consistent adoption trajectories. The best-fitting curve is selected based on the highest $R^2$ value, conditional on exceeding 0.9. The AUC of each geographic entity ($A_i$) is derived from its best-fitting curve. 

\smallskip

(2) \textit{AUC of mean cumulative Adoption} ($A_m$): For each time point ($t$), the regional mean of adoption intensities is computed. This is done based on the values of all entities included in the region and participating in the process. The best-fitting cumulative curve for the means' time series is identified ($C_m$), and AUC is computed ($A_m$). Subsequently, the AUC of each geographic entity ($A_i$) is normalized relative to the regional means AUC as $\tilde{A}_{i} = (A_i / A_m) \cdot 100$. By definition, this normalization sets the normalized AUC of the regional means ($\tilde{A}_{m}$) to 100.

\smallskip

(3) \textit{The feedback mechanism:} ATI provides a unique solution for capturing temporal adoption dynamics of each geographic entity by responding to its deviations from the regional mean trend. $\tilde{A}_{i}$ is adjusted through this mechanism, which operates in one of two opposing directions: a penalty or a reward. A penalty is imposed on entities whose overall AUC exceeds the regional mean ($\tilde{A}_i > 100$), yet falls below the regional mean curve during specific periods.
A reward is granted to entities whose AUC is below the regional mean ($\tilde{A}_i < 100$), but that exceeded the mean curve during specific periods. In both cases, the feedback quantifies the duration of time the entity remained below or above the regional mean curve, respectively, with the resulting adjustment being proportional to that duration. The adjustment is thus directly time-proportional. 
This mechanism facilitates the distinction between entities with strong and stable adoption patterns and those that are strong but volatile, by penalizing the latter for their instability; conversely, it rewards entities that manage to close the gap with the regional mean, albeit temporarily, thus distinguishing them from those that are consistently weak. Entities that maintain a consistent trend throughout the entire period will not receive any penalty/reward.
The feedback adjustment for an entity is determined by the total duration of time intervals between consecutive intersection points $j$ (each corresponding to a time value $t_j$) where the entity’s curve $C_i$ intersects with the regional mean curve $C_m$. These time intervals represent the entity's deviations from its overall adoption trend. The angle of each deviation determines whether the corresponding time interval is subject to feedback adjustment, as follows: For each intersection point $j$, the sine of the angle $\alpha_j$ formed immediately after the intersection is computed. This value indicates whether the entity’s curve $C_i$ is increasing or decreasing relative to the regional mean curve $C_m$. A positive sine value of $\alpha_j$ indicates that $C_i$ rises above $C_m$ after $t_j$, whereas a negative sine value indicates that $C_i$ falls below $C_m$.
It is important to note that the feedback mechanism is based on the length of deviation intervals, measured along the time axis, rather than on the area under the curve of these intervals, as the latter is already captured in $\tilde{A}_i$.

ATI adjusted the $\tilde{A}_i$ value only for entities that intersect with $C_m$, and the calculation is carried out in three steps. \textit{First}, the feedback factor $S_{i,j}$ is computed for each intersection point (Eq. \ref{S}). This factor represents the combination of the $sin(\alpha_j)$ and the difference between the $\tilde{A}_{i}$ and $\tilde{A}_{m}=100$ ($\Delta_i$). $S_{i,j}$ incorporates the sign function ($\text{sgn}(x)$), which returns $1$ for $x > 0$, $-1$ for $x < 0$, and $0$ otherwise. 
\vspace{0.2em}
\begin{equation}
\centering
\resizebox{0.5\hsize}{!}{$
S_{i,j} = \frac{1 - \text{sgn}(\sin\alpha_{i,j}) \cdot \text{sgn}(\Delta_i)}{2}
$}
\label{S}
\end{equation}

\textit{Second}, the feedback corresponding to each deviation is calculated. The feedback for the time interval between the initial time point of the timeline and the first intersection point ($F_{i,t_{k=1}}$) is determined by Eq.\ref{f1}. Subsequently, the feedback for each time interval between consecutive intersection points ($F_{i,j}, \dots, F_{i,n}$) is determined by Eq.\ref{fj}.
\vspace{-0.1em}
\begin{equation}
\resizebox{0.52\hsize}{!}{$
F_{i, t_{k=1}} = (t_{i,j} - t_{k=1}) \cdot\left(1-S_{i,j}\right)
$},
\label{f1}
\end{equation}
\vspace{-1.5em}
\begin{equation}
\resizebox{0.8\hsize}{!}{$
\begin{aligned}
    F_{i,j} &= \left( t_{i,j+1} - t_{i,j} \right) \cdot S_{i,j}\\
    &\quad \text{for} \quad j = 1, j+1, j+2, \dots, n, \quad t_{i,n+1}=t_{k=T,}
\end{aligned}
$}
\label{fj}
\vspace{2em}
\end{equation}

where $t_{k=1}$ is the first time point of the timeline. $t_{i,j}$ is the time of the first intersection point, $t_{i,j+1}$ is the next intersection point, etc. Note that for the last intersection point, $t_{i,n+1}$ corresponds to the final point on the timeline ($t_{k=T}$).

\textit{Lastly}, the index is calculated by Eq.\ref{index}:  
\vspace{-0.1em}
\begin{equation}
ATI_i =  \tilde{A}_i - (\dfrac{F_{i,t_{k=1}}+\sum_{j=1}^{n} F_{i,j}}{t_{k=T}- t_{k=1}}) \cdot \Delta_i, 
\label{index}
\vspace{-0.5em}
\end{equation}
where $ATI_i$ is the index result for unit $i$, the numerator of the expression in parentheses denotes the total duration of deviation periods, and the full expression captures the proportion of these periods relative to the total timeline. This expression, multiplied by $\Delta_i$, is the overall penalty or reward. Notice that a negative $\Delta_i$ will result in a reward, while a positive $\Delta_i$ will result in a penalty. For entities that do not have any intersection points $ATI_i = \tilde{A}_i$.\\

\begin{table}[h]
\centering
\renewcommand{\arraystretch}{0.99} 
\setlength{\tabcolsep}{5pt} 
\captionsetup{width=\columnwidth}
\caption{\textbf{Pseudo-code of adoption over time index (ATI) computing algorithm.} The best-fitting curve is calculated from a set of candidate curves, which includes Logistic, Gompertz, Bass, Generalized Richards, Cumulative Normal, Exponential, Bertalanffy, Polynomial curves of degree $d$ (where $d \in \{2, \ldots, T-1\}$), and Linear curve. $AI$ - adoption Index; $t$ - time.}
\begin{tabular}{p{0.999\columnwidth}}\raggedright 
\rule{\columnwidth}{1pt} 
\textbf{Input:} \\
$t:= \{t_k \mid k \in \{1, \dots, T\}\}$\\
$AI:= \{AI_{i,t} \mid i \in \{1, \dots, N\}, t \in \{t_{k=1}, \dots, t_{k=T}\}\}$\\
\vspace{0.1cm}
\textbf{Output:} \\[1pt]
$ATI:= \{ATI_i \mid i \in \{1, \dots, N\}\}$ \\
\vspace{-0.1cm}   
\rule{\columnwidth}{0.4pt} \\[3pt]
1.  For $t_k \in \{t_{k=1}, \dots, t_{k=T}\}$: \ $(1/N) \cdot \sum_{i=1}^{N} AI_i \rightarrow AI_{m,t}$  \\
\vspace{0.15cm}
2. Fit best curve: \\
\hspace{0.3cm} $\arg\max \left(R^2(C, AI_{m,t}) \mid C \in \text{curves set}\right) \rightarrow C_m$ \\
\vspace{0.15cm}
3. Compute AUC of $C_m$: \ $\int C_m \rightarrow  A_m  $ \\
\vspace{0.15cm}
4. Normalized $A_m$: $A_m \mapsto \tilde{A}_m = 100$\\   
\vspace{0.15cm}
5. For $i \in \{1, \dots, N\}$: \\
\vspace{0.12cm}
\hspace{0.4cm}- Fit best curve: \\
\hspace{0.5cm} $\arg\max \left( R^2(C, AI_{i,t}) \mid C \in \text{curves set} \right) \rightarrow C_i$ \\
\vspace{0.12cm}
\hspace{0.4cm}- Compute AUC of $C_i$: \ $\int C_i \rightarrow A_i$ \\
\vspace{0.12cm}
\hspace{0.4cm}- Normalize $A_i$: \ $\left(A_i / A_m\right) \cdot 100 \rightarrow \tilde{A}_{i}$ \\
\vspace{0.12cm}
\hspace{0.4cm}- Find time values ($t$) of intersection points ($j$):\\ 
\hspace{0.65cm}$\{t \mid C_i(t) = C_m(t), \, j \in \{1, \dots, n\}\} \rightarrow t_{i,j} $ \\
\vspace{0.12cm}
\hspace{0.4cm}- If $n \geq 1$: \\
\vspace{0.10cm}
\hspace{0.9cm}For $j \in \{1, \dots, n\}$:\\
\vspace{0.10cm}
\hspace{1.4cm}Compute $\arctan(C'_i(t_{i,j}) - C'_m(t_{i,j})) \rightarrow \alpha_{i,j}$ \\
\vspace{0.10cm}
\hspace{1.4cm}Compute $S_{i,j}$ \,(\text{\scriptsize Eq.\ref{S}})\\
\vspace{0.10cm}
\hspace{1.4cm}Compute $F_{i, t_{k=1}}$ \,(\text{\scriptsize Eq.\ref{f1}})\\ 
\vspace{0.10cm}
\hspace{1.4cm}Compute $F_{i,j}$ \,(\text{\scriptsize Eq.\ref{fj}})\\
\vspace{0.12cm}
\hspace{0.9cm}Compute $ATI_i$  \,(\text{\scriptsize Eq.\ref{index}})\\
\vspace{0.12cm}
\hspace{0.6cm}Else: $ATI_i = \tilde{A}_i$ \\
\vspace{0.2cm}
6. Return: $\{ATI_i \mid i \in \{1, \dots, N\}\}$ \\
\rule{\columnwidth}{1pt} 
\end{tabular}   
\label{algorithm}
\end{table}

\begin{figure}[]
    \centering
    \setlength{\fboxsep}{0.5pt} 
    \setlength{\fboxrule}{0.5pt} 
    \fbox{\resizebox{0.87\columnwidth}{!}{\includegraphics{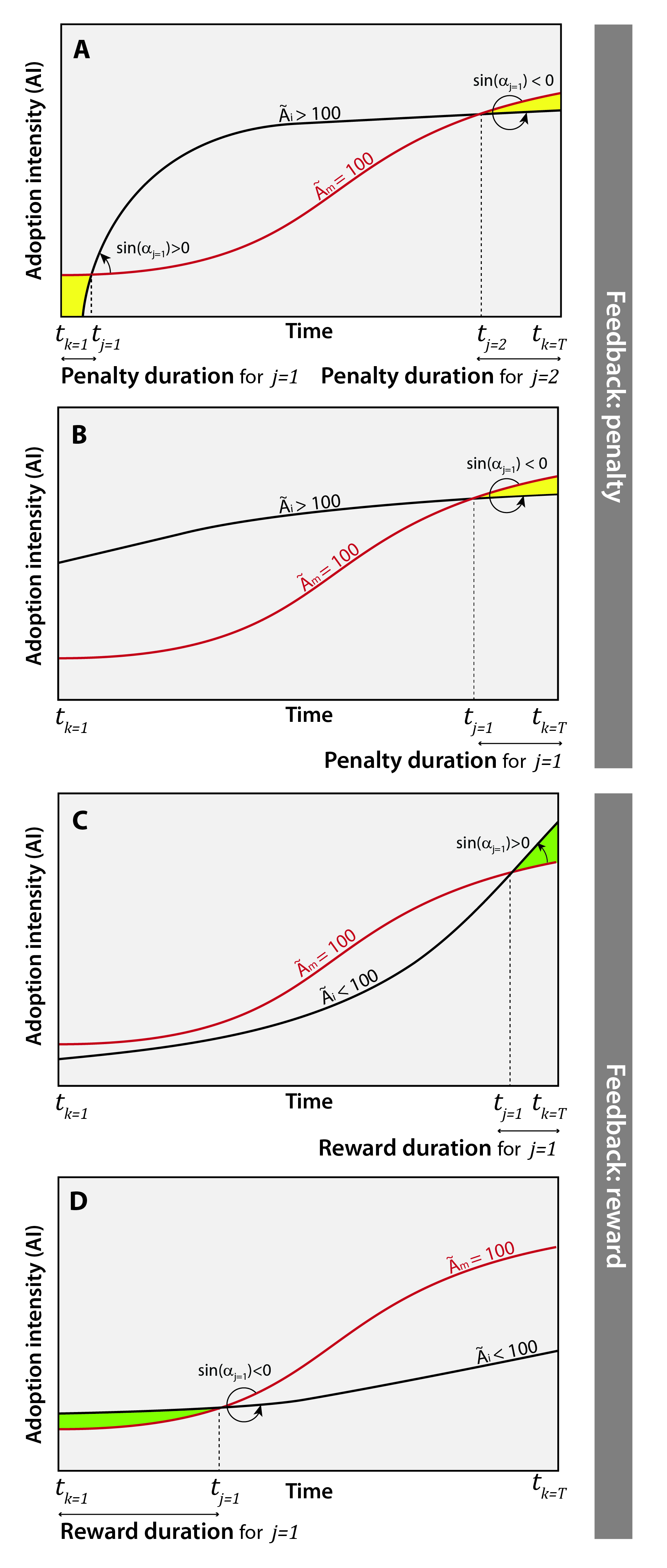}}}
    \caption{\textbf{Illustrative Examples of Feedback Scenarios.} The feedback is determined by the normalized value of $\tilde{A}_{i}$, which represents the entity’s AI relative to the regional mean. The time periods over which rewards or penalties are calculated depend on the type of feedback and on the angle formed after each intersection point with the regional mean curve. Note: The AI axis range is context-dependent and determined by the chosen AI metric and the adoption magnitudes in the case examined.} 
    \label{feedback}
\end{figure}

To demonstrate the feedback mechanism, four basic scenarios are outlined in Fig.\ref{feedback} and summarized briefly below. While simplified, these scenarios serve as illustrative examples of the feedback mechanism. Naturally, real-world cases may be more complex and incorporate multiple intersection points. 

\begin{itemize} [leftmargin=1em, itemsep=0pt]
    \item An entity with $\tilde{A}_i > 100$ and $sin(\alpha_j) > 0$ (Fig. \ref{feedback}A): \\
    Although this entity stands out in terms of adoption intensity, as indicated by its $\tilde{A}_{i}$, the presence of an intersection point followed by a positive angle suggests that prior to this intersection, its adoption levels were below the mean. Therefore, despite its high overall adoption intensity, it cannot be regarded as a definitive leader, due to its earlier under-performance and delayed entry. If a second intersection is observed, it indicates that after surpassing $C_m$, it experienced another deceleration. The feedback adjustment accounts for its delayed rise and/or regression periods and, therefore, lowers its leadership score. 
    \item An entity with $\tilde{A}_{i} > 100$ and $sin(\alpha_j) < 0$ (Fig. \ref{feedback}B): \\
    A negative sine value of the first intersection point indicates that $C_i$ was above $C_m$ early in the timeline but subsequently dropped below the regional mean until the next intersection point. In this case, ATI penalizes the entity proportionally to the duration it remained below the mean.     
    \item An entity with $\tilde{A}_{i} < 100$ and $sin(\alpha_j)> 0$ (Fig. \ref{feedback}C):\\
    This entity clearly falls into a weak adoption segment, but since it remains above $C_m$ for a while, ATI rewards it accordingly. This compensation quantifies its potential to exceed the regional mean curve despite its overall low performance.
    \item An entity with $\tilde{A}_{i} < 100$ and $sin(\alpha_j) < 0$ (Fig. \ref{feedback}D):\\
    Negative trajectory relative to $C_m$ implies that prior to the intersection point, the unit was above the regional mean. Therefore, ATI compensates it proportionally to this growth phase.
\end{itemize}

Note: The AI axis range is context-dependent and is determined by the chosen AI metric and the adoption magnitudes in the case examined.

Several important observations about the feedback mechanism should be noted: First, the feedback is applied solely to the difference between $\tilde{A}_{i}$ and $\tilde{A}_{m}$. Therefore, penalty/reward cannot reduce/increase ATI values beyond this difference, and the entity will always retain its position relative to the regional mean (either above or below). Second, the feedback mechanism intensifies as $\tilde{A}_{i}$ diverges from $\tilde{A}_{m}$. Namely, entities that exhibit large deviations from the regional average, combined with shifts in their trajectories over time, are subject to greater penalties or rewards. This refinement ensures that only persistently leading or lagging behavior will be classified as such. 

\subsubsection{Entry time}\label{sec: time} Differentiating between the adoption types outlined in Rogers' Diffusion of Innovations theory depends largely on the timing of an entity’s entry into the adoption process. This feature is widely used to assess the nature of adoption and to compare its extent across different entities. In PV studies, this feature is commonly defined as the point in time when solar energy accounts for a specified share of the total electricity system, as measured by production, capacity, or supply. However, the thresholds used vary widely across studies, typically ranging between 1\% and 10\% \citep{cherp2021national, bento2016measuring}. Since in many off-grid communities, such as in our case study, there is no systematic estimation of PV capacity or collection of consumption data, we define a relative metric that captures the point in time at which an entity reaches a designated threshold of adoption intensity, expressed as a fraction of the regional mean intensity. The methodological rationale and value of the threshold used in our case study are provided in Section \ref{sec:feature_extraction}.

\subsubsection{Latest Adoption Intensity (LAI)} \label{sec: LAI}
While ATI represents adoption intensities throughout the entire timeline, it is crucial to measure the most recent adoption intensity, i.e., intensity at the last recorded time point (LAI). This feature sheds light on the current state of adoption in the investigated entity and allows comparisons between its overall adoption process and its latest outcome. LAI is normalized relative to the adoption intensity of the regional mean at the same time point, with the latter assigned a reference value of 100. This normalization allows for assessing the entity's relative performance and facilitates comparison with ATI, which is inherently normalized in the same manner.

\subsubsection{Latest trajectory} \label{sec:trajetory}
ATI accounts for trend shifts occurring along the timeline to refine the quantification of the entity's adoption intensity and differentiate between adoption dynamics. Particular importance is given to the entity's most recent trend relative to the regional mean, as it provides insight into its future adoption progression. The latest trajectory is a ternary ordinal variable that represents the final slope direction of the entity’s curve relative to the regional mean curve. This feature has three possible values: \textit{Uphill} assigned when the sine of $\alpha$ after the last intersection point $t_n$ is positive, indicating that the entity’s adoption intensity is increasing at a faster rate than the regional mean; \textit{Downhill} assigned when the sine of $\alpha$ after the last intersection point $t_n$ is negative, indicating that the change in adoption intensities rate is slower than the regional mean; and \textit{Stable} assigned to entities whose curve has no intersection points with $C_m$, meaning its relative trend remains stable along a timeline, either consistently above or below the regional mean. A null value is assigned to entities with a zero ATI value. 

\subsection{Profiling} \label{sec: profiling}
The four key features serve as the basis for creating a profile for each geographic entity. To reduce the broad range of possibilities resulting from the continuous features - ATI, entry time, and LAI - their scale is categorized into an ordinal scale. ATI and LAI are converted into three ordinal tiers, defined as follows: The mean feature value $\pm 0.44\sigma$ is classified as \textit{medium}, while values below or above this range are classified as \textit{low} and \textit{high}, respectively. This classification is based on Rogers' Diffusion of Innovations theory \citep{rogers2003diffusion}, which defines adoption type by using the standard normal distribution. Following Rogers’ definition, our classification ensures that the medium tier includes 34\% of the normal distribution, corresponding to the combined proportion of half of the early majority and half of the late majority. The two symmetric adoption categories flank the average adoption level. A naught is assigned to entities whose feature value is equal to zero. 
Entry time is also classified into three ordinal levels: early, middle, and late. The middle value serves as a temporal benchmark, defined as the mean of all entities’ entry times, that is, the average time at which the adoption threshold is first met across the region. Entities whose entry times fall within a range of $\pm 0.44\sigma$ around this benchmark are classified as \textit{medium}, whereas earlier and later entry times are classified as \textit{early} and \textit{late}, respectively. A null value is assigned to units that did not reach the required threshold value throughout the studied timeline.

\begin{table*}[h]
    \centering
    \captionsetup{width=\textwidth}
    \caption{\textbf{Criteria for the typology of adoption paths.} The continuous features were assigned to ordinal scales. Combinations of these feature values enable distinct definitions of adoption behaviors over time. A general scheme of these criteria is presented here.}
    \renewcommand{\arraystretch}{1.2} 
    \setlength{\tabcolsep}{5pt} 
    \begin{tabular}{m{3cm}m{2cm}m{3.05cm}m{3cm}m{4cm}}
        \Xhline{1.2pt}
        \textbf{Adoption Path} & \textbf{ATI} & \textbf{Entry Time} & \textbf{Latest Trajectory} & \textbf{Latest adoption intensity} \\
        \hline
        \textbf{Leading} & High & \hspace{5mm}Early & \hspace{6mm}Stable & \hspace{12mm}High \\
       
        \textbf{Accelerating} & High & \hspace{5mm}Early & \hspace{6mm}Uphill & \hspace{12mm}High \\
       
        \textbf{Decelerating} & High & \hspace{5mm}Early & \hspace{6mm}Downhill & \hspace{12mm}Medium/Low \\
       
        \textbf{Leaping} & Low/Medium & \hspace{5mm}Late & \hspace{6mm}Uphill & \hspace{12mm}High \\
       
        \textbf{Moderate} & Medium & \hspace{5mm}Early/Middle & \hspace{6mm}Stable/Uphill & \hspace{12mm}Medium \\
        
        \textbf{Declining moderate} & Medium & \hspace{5mm}Early/Middle & \hspace{6mm}Downhill & \hspace{12mm}Low \\
       
        \textbf{Lagging} & Low & \hspace{5mm}Early/Middle/Late & \hspace{6mm}Stable/Downhill & \hspace{12mm}Low \\
        
        \textbf{Non-Adopting} & Zero & \hspace{5mm}None & \hspace{6mm}None & \hspace{12mm}Zero \\
        \Xhline{1.2pt}
    \end{tabular}
   
    \label{Criteria}
\end{table*}

A profile for each entity is constructed as a combination of its ordinal values across the four key features. In total, 144 theoretically possible combinations emerge from this full crossing of ordinal values. However, some combinations are infeasible given the cumulative nature of the innovation adoption process. For instance, a profile with late entry time, low LAI, negative trajectory, and high ATI is not possible, as a high ATI inherently requires sustained accumulation over time. Other examples include a combination of low ATI, late entry, negative trajectory, and high LAI, or high ATI with early entry time, positive trajectory, and low LAI. Forty-five such infeasible combinations were filtered out, leaving 99 profiles consistent with the domain-specific context and logic of the research field. These profiles served as the basis for the next stage of typology creation.

\subsection{Typology of adoption paths} \label{sec: typology}
\begin{figure*}[]
    \centering
    \setlength{\fboxsep}{0.5pt} 
    \setlength{\fboxrule}{0.5pt}
     \fbox{\resizebox{0.9\textwidth}{!}{\includegraphics{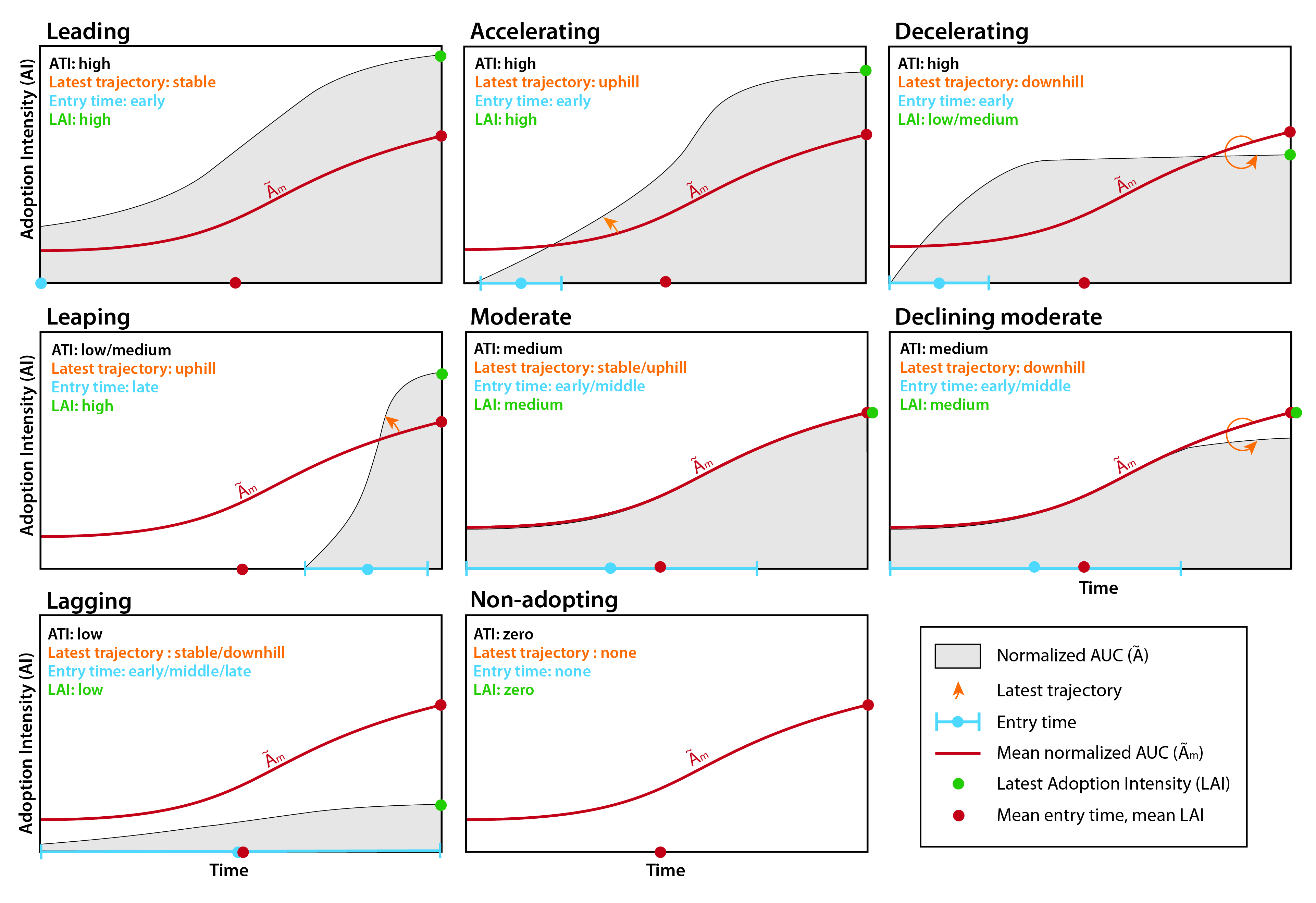}}}  
    \caption{\textbf{Typology of adoption paths}. Eight conceptual adoption paths are illustrated, each representing a distinct adoption dynamic. The paths are determined based on a set of criteria (Table \ref{Criteria}) derived from ordinal categories of four key features: ATI, Entry time, Latest trajectory, and LAI. Note: The AI axis range is context-dependent and determined by the chosen AI metric and the adoption magnitudes in the case examined.}
    \label{theoretical_paths}
\end{figure*}

An analysis of the possible profiles was conducted to identify adoption behaviors over time and to define a typology based on an \textit{if-then criteria} system (Table \ref{Criteria}). Eight distinct adoption behavior paths were identified and termed \textit{adoption paths}. An illustration of the different paths throughout the timeline, along with their key feature values, is presented in Fig. \ref{theoretical_paths}. The overview below provides a concise definition of their adoption attributes:
\begin{enumerate} [leftmargin=1em, itemsep=0pt]
\item \textit{Leading path:} Entities that exhibit high adoption intensity from the very beginning of the timeline and maintain elevated levels throughout. These entities do not intersect with the regional mean curve; their trajectory remains stable, with adoption intensity consistently high through the end of the timeline.
\item\textit{Accelerating path:} Entities that enter the adoption process slightly later than the leading path, but exhibit a clear uphill trajectory, rapidly reaching high adoption intensities. These entities intersect the regional mean curve only once, early in the timeline, and maintain high adoption levels consistently thereafter.
\item\textit{Decelerating path:} Similar to the Accelerating Path, these entities enter the adoption process relatively early and initially achieve high adoption intensities. However, at a later stage in the timeline, a slowdown or stagnation occurs, ultimately resulting in a downhill trajectory, and adoption levels fall below the regional mean by the end of the timeline.
\item\textit{The leaping path:} These entities are characterized by a late entry into the adoption process, followed by a sharp increase in adoption intensity and a sustained uphill trajectory, ultimately reaching high adoption levels by the end of the timeline.
\item\textit{Moderate path:} Entities follow a trajectory similar to the regional mean, with varying entry times. Intersection points with the mean curve may or may not occur, and their final adoption intensity typically aligns with the regional intensities.
\item\textit{Declining moderate path:} These entities exhibit adoption intensities similar to the regional mean. However, their latest trajectory is downhill, reflecting a decline or stagnation in adoption towards the end of the timeline, with final adoption levels falling below the regional average.
\item\textit{Lagging path:} These entities are characterized by consistently low adoption intensities, entry times distributed across the timeline, and adoption rates that remain below the regional mean throughout the entire period.
\item\textit{Non-adopter path:} Entities that consistently remain outside the adoption process throughout the entire timeline, exhibiting no measurable adoption activity.

\end{enumerate}

\subsection{Application of the proposed framework}
\begin{figure*}[]
    \centering
    \setlength{\fboxsep}{0.5pt} 
    \setlength{\fboxrule}{0.5pt}
    \fbox{\resizebox{0.9\textwidth}{!}{\includegraphics{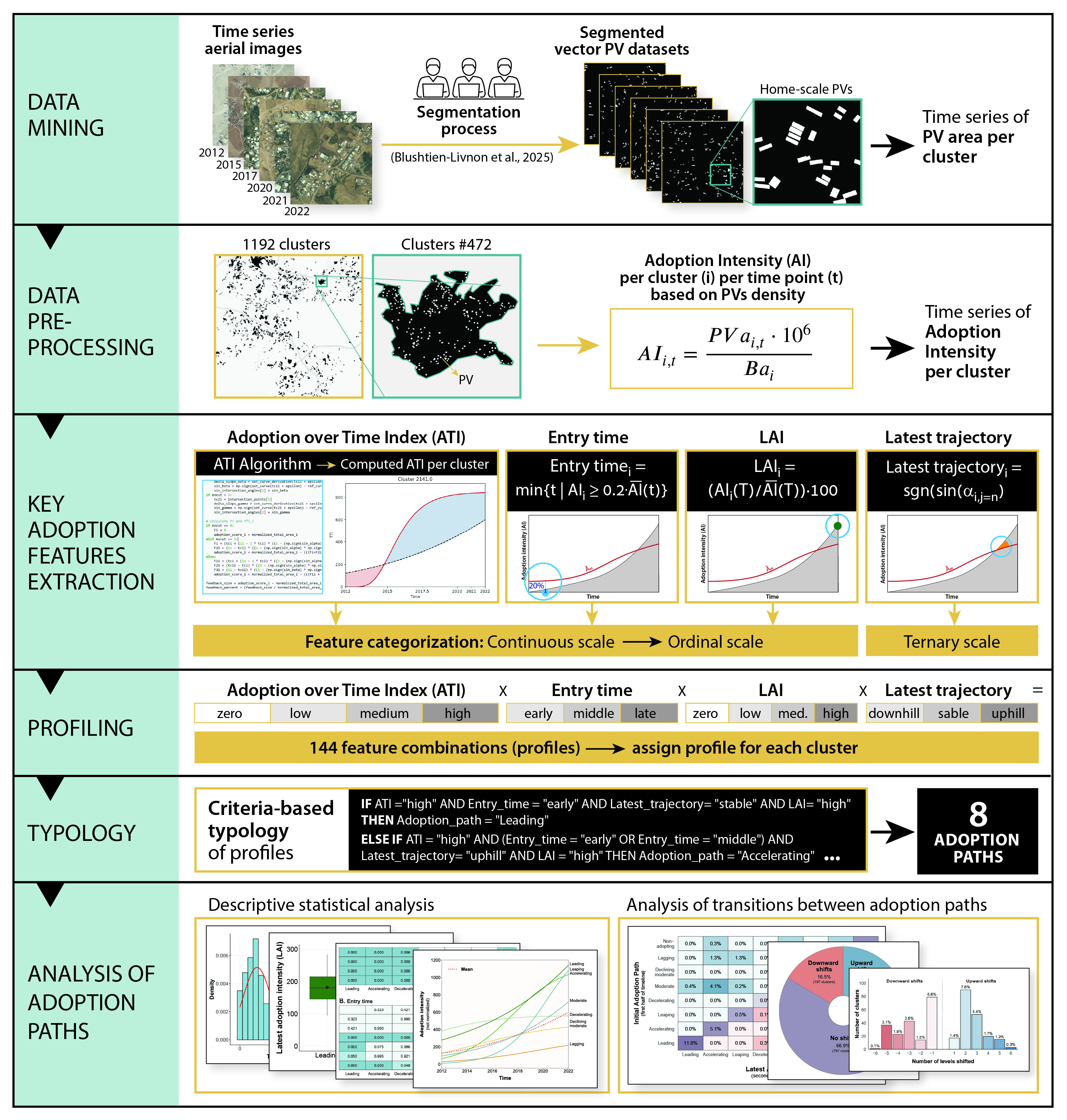}}}
    \caption{\textbf{Analytical Framework for the Typology of Adoption Paths.} The research progresses from PV monitoring over time to density computation, quantification of four key adoption-defining features, and their integration into diverse profiles. These profiles serve as the typological foundation for distinct adoption behaviors. Finally, a case study provides empirical evidence for the developed framework.} 
    \label{framework}
\end{figure*}

To validate the proposed analytical framework of adoption paths in a real-world setting, we applied it to a case study in Israel. Fig. \ref{framework} illustrates the framework implementation steps, culminating in an analysis of PV adoption dynamics within the investigated region and its characteristics based on adoption path typology.

\subsubsection{Geographical context}
\begin{figure}[]
    \flushleft
    \setlength{\fboxsep}{0.5pt} 
    \setlength{\fboxrule}{0.5pt} 
    \fbox{\includegraphics[width=\columnwidth]{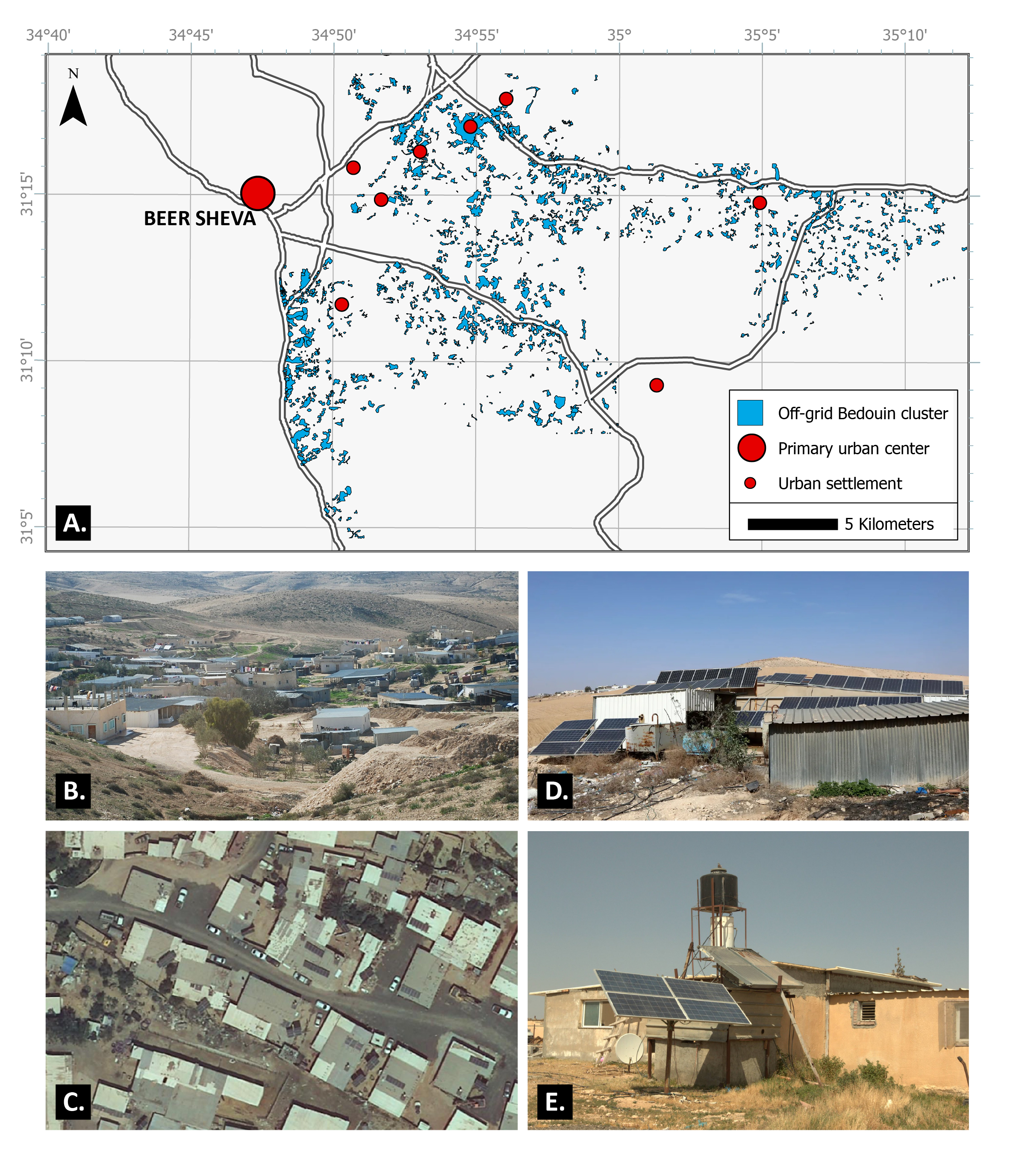}}
    \caption{\textbf{Off-grid settlement clusters and typical home-scale PV deployment in the study area.}} A: Scattered distribution of Bedouin clusters in southern Israel, spanning \textasciitilde{1,700 km²}. The clusters vary in area size. B: Ground-level view of a cluster \citep{neukolnBedouinVillages}. C: Aerial view illustrating the unique spatial organization of a Bedouin cluster, characterized by irregularly arranged structures that lack conventional planning features. The dwellings primarily consist of tin structures with a few stone buildings. PVs are installed both on rooftops and on the ground. D-E: Typical PV installations within a cluster, highlighting the informal and decentralized nature of solar energy deployment. \footnotesize(Courtesy: D - Miki Kratsman; E - Eliyahu Hershkovitz, Haaretz).   
    \label{study_area}
    \end{figure}

As a case study, we selected a vast region in the Negev Desert, Israel, inhabited by \textasciitilde{100,000} Bedouins, an Arab minority group. The inhabitants of this region reside in 1,192 widely dispersed (Fig. \ref{study_area} A-B), unauthorized off-grid settlements, referred to as clusters, which are characterized by low socioeconomic status and severe energy deprivation \citep{teschner2024energy}. Until a decade and a half ago, energy production in this population relied primarily on expensive and polluting diesel-powered generators. However, since the early 2010s, inhabitants of these clusters have increasingly adopted small-scale PV systems as a more affordable and efficient solution for electricity supply \citep{shapira2021energy} (Fig. \ref{study_area} C-E). This shift reflects broader global trends in off-grid communities, with total installed PV electricity capacity in off-grid communities worldwide growing fourfold between 2014 and 2023 \citep{IRENA2024OffGrid}. Preliminary data on the Bedouin off-grid communities indicate that their PV adoption process combines rapid growth with significant spatial variability, mirroring patterns also observed in other off-grid societies (e.g., \citep{aarakit2021adoption, akter2021off, saha2025empowering}. These parallels underscore the relevance of the Bedouin case as an insightful example for understanding PV adoption processes in off-grid marginalized communities and for informing policy strategies aimed at promoting adoption in similar contexts.
   
\subsubsection{Data mining}
Since PV installations in Bedouin settlements are carried out informally, there is no official registry of residential PVs used by the studied population. Consequently, remote sensing imagery and a segmentation-based data mining process were employed to identify PVs and quantify their area. A series of six RGB aerial images from the years 2012, 2015, 2017, 2020, 2021, and 2022 served as a basis for PV data extraction. The spatial resolution of the images ranged from 0.25 meters (2012-2015) to 0.14 meters (2017–2022). The segmentation process was implemented by an annotation team with expertise in interpreting and digitizing remote sensing imagery and was carried out using ArcGIS Pro. The process followed an established segmentation protocol \citep{blushtein2025performance}, which included two rounds of performance validation by an independent annotation team, and an estimated accuracy of 0.885 (F1 score). The final polygonal vector layers served as raw data for the adoption intensity (AI) calculation. 

\subsubsection{Data pre-processing: Creating Adoption Intensity (AI) time series} \label{sec: AI}
Since the studied geographical entities, household clusters, vary in size, the cumulative segmented area of PVs, in each cluster, must be normalized to population size. The clusters are not recognized as legal by state authorities, and therefore, no formal demographic data is available. As a proxy, we normalized PV area to the built-up area within the clusters. Data on cluster boundaries was obtained from the Authority for Development and Settlement of the Bedouin in the Negev. Cluster boundaries were delineated based on the extent of built-up areas within each cluster. Since the clusters consist exclusively of single-story tin structures with relatively homogeneous building densities, built-up area served as a proxy for population size.

Adoption Intensity (AI) for each cluster, at each time point, was calculated as follows:
\begin{equation}
 AI_{i,t} = \frac{PVa_{i,t} \cdot 10^6}{Ba_i},
\label{AI}
\end{equation}

\noindent where $PVa_{i,t}$ is the total PV area in cluster $i$, at time point $t$, and $Ba_i$ is the built-up area of cluster $i$. The constant $10^6$ is used to avoid the use of small floating-point numbers. 

\subsubsection{Key adoption features extraction} \label{sec:feature_extraction}
Extraction of three key features - ATI, LAI, and Latest trajectory - was conducted as described in Sections \ref{sec: ATI}, \ref{sec: LAI}, and \ref{sec: trajectory}, respectively. To extract the entry time feature (Section~\ref{sec: time}), which represents the point in time at which each entity reaches a threshold fraction of the regional mean adoption intensity, a decision had to be made regarding the appropriate threshold value. To guide this decision, the distribution of entry times was examined across various threshold fractions of the regional mean, in 10\% increments ranging from 10\% to 60\%. This was done to: (1) Identify threshold fraction at which the distribution of entry time values most closely approximates a normal distribution, ensuring that the studied entities are distributed as symmetrically as feasible around their central tendencies; (2) Ensure that the majority of the studied entities reach this threshold fraction at some point within the examined timeline; and (3) Identify a threshold threshold value with sufficiently large variance. A threshold value that meets these criteria would serve as the most effective measure for distinguishing between different adoption dynamics. Our data indicated that 20\% of the regional mean is the most appropriate threshold value. Compared with other candidate thresholds, this value yielded the distribution with skewness and kurtosis values closest to a normal distribution (see Fig. \ref{variables_stat}B), had the highest SD, and was reached by approximately 92\% of the settlement clusters at some point along the timeline.

\subsubsection{Profiling and typology}
Defining a data-driven profile for each cluster in the study area, and assigning it to the corresponding adoption path according to the criteria-based typology outlined above was straightforward and carried out following the procedure detailed in Sections \ref{sec: profiling} and \ref{sec: typology}, respectively.

\subsubsection{Analysis of adoption paths}
Application of adoption path typology was conducted in two levels: (1) Statistical analysis of adoption paths: We conducted a descriptive statistical analysis of the data-driven adoption paths, which includes statistical moments evaluation of key variables, an examination of adoption paths distribution, an assessment of quantitative distinctions between paths using ANOVA and post-hoc tests, calculation of median curves of the paths and an evaluation of their alignment with the conceptual adoption paths presented in section \ref{sec: typology} and Fig \ref{theoretical_paths}. (2) Temporal transitions between adoption paths: To examine transitions in adoption behavior, the timeline was divided into two equal parts, and the midpoint was set to 2017. Each of the key features was calculated for each timeline half accordingly: ATI was computed based on the curve's area of the respective half, considering only intersection points occurring within that period. The LAI for the first half of the timeline was determined based on adoption intensity in 2017 and normalized to the regional mean of that year, and the LAI for the second half corresponds to its 2022 normalized value. The latest trajectory was derived from the last intersection point within each half. In contrast, entry time values remained unchanged, as they must be referenced to the full timeline, and any comparison of time points between the different halves would be meaningless. Next, clusters were classified into different paths based on their adoption characteristics in each part of the timeline, and the transitions between the two halves were analyzed using a transition matrix, or a heatmap. Lastly, since the adoption paths can be ordered along an ordinal scale, according to their progression, we defined downward transitions as shifts from a higher to a lower tier, and upward transitions as shifts from a lower to a higher tier within this hierarchy. The paths ordering from lowest to highest tier of adoption were defined as follows: The Non-adopting path represents the lowest tier, followed by Lagging, Declining Moderate, Moderate, Decelerating, Leaping, Accelerating, and Leading, which represents the highest tier. Organizing the paths in this ordinal structure enabled us to quantify the magnitude of transitions occurring between the two halves of the timeline.

\section{Results}
\subsection{ATI vs. \~{A}}
For each cluster's time series, a set of cumulative curves (Section \ref{sec: ATI}) was fitted. The curve with the highest R², provided it exceeded 0.9, was then selected and used as the basis for calculating the cluster's AUC. ATI integrates the quantification of both cumulative adoption intensities over time and temporal dynamics. These dynamics are captured through a feedback mechanism that rewards/penalizes entities based on their deviation from the regional mean and the consistency of their adoption trajectory. Thus, ATI differs from $\tilde{A}$, which reflects only AUC, lacking consideration of trend dynamics. To highlight the contribution of the reward/penalty feedback mechanism to the basic AUC calculation, Table \ref{comparison} compares ATI vs. $\tilde{A}$ across key statistical tests and measures. The Wilcoxon signed-rank test, applied to evaluate if systematic differences between $\tilde{A}$ and ATI exist, revealed a statistically significant result (p-value < 0.0000). This test was selected due to the paired nature of the metrics for each cluster and their deviation from normal distribution, confirmed by the Shapiro-Wilk test (p-value < 0.05). Comparing statistical moments of the metrics reveals several key insights: (1) Although their mean values are similar, ATI exhibits lower overall variability, as reflected in its smaller standard deviation and reduced gap between the mean and the median. This suggests that ATI draws values closer to the center of the distribution, consolidating those clustered around the regional average; (2) ATI displays higher kurtosis and increased positive skewness, indicating that it accentuates deviations at the distribution’s extremes. Consequently, ATI simultaneously reduces dispersion among average-range cases while enhancing the distinction of cases with exceptional adoption intensities, making it a more discriminative and informative metric for capturing differences in adoption paths than $\tilde{A}$.

\begin{table}[h]
    \centering
    \caption{Comparison of Key Statistical Moments for \~{A} (AUC of Adoption Intensity Over Time) and ATI}
    \begin{tabular}{m{0.6cm}cccm{0.7cm}c}
        \toprule
        & Mean & Median & Std Dev & Kurtosis & Skewness \\
        \midrule
        \~{A}   & 100.000  & 81.7942  & 85.2748  & 7.1998  & 1.7290  \\
        ATI & 96.8102  & 88.7374  & 77.3249  & 8.6641  & 1.9777  \\
        \midrule
        \multicolumn{6}{l}{\textbf{Wilcoxon Signed-Rank test:} W = 47096, p-value < 0.0000} \\
        \bottomrule
    \end{tabular}

    \label{comparison}
\end{table}

\subsection{Descriptive statistics of key adoption features}
\begin{figure*}[]
    \centering
    \setlength{\fboxsep}{0.5pt} 
    \setlength{\fboxrule}{0.5pt} 
    \fbox{\resizebox{0.9\textwidth}{!}{\includegraphics{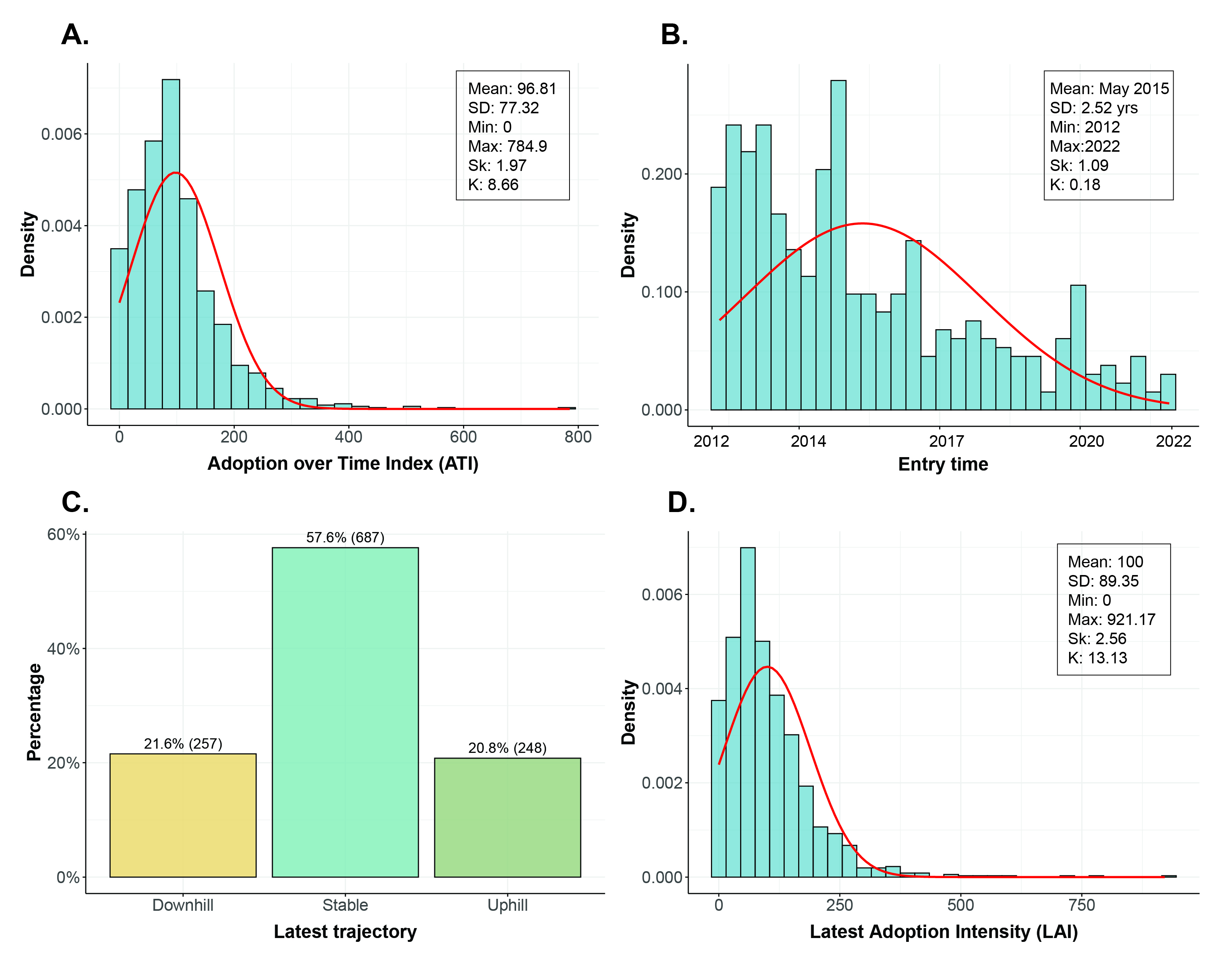}}}
    \caption{\textbf{Descriptive statistics of key adoption features:} A. ATI: An observed right-skewed distribution indicates that most clusters have low ATI, while a few exceed the regional mean ($ATI_i$>200). High skewness and kurtosis indicate outliers with exceptionally high adoption levels. B. Entry time: The moderately skewed distribution with low kurtosis indicates that entry times are widely dispersed. The distribution peaks early in the timeline and again in 2014–2016, followed by a gradual decline. C. Latest trajectory: A trend shift relative to the regional mean was observed in 42.4\% of the clusters, indicating that the adoption process evolves over time. The balance between up- and downhill trajectories suggests heterogeneity in long-term adoption patterns. D. The right-skewed distribution of LAI indicates that most clusters still fall below the mean. However, compared with ATI, outlier clusters exhibit higher adoption intensities at the latest time point. Overall, the results highlight substantial variability in adoption across clusters and trend shifts over time.}    
    \label{variables_stat}
\end{figure*}

Fig. \ref{variables_stat} provides a statistical overview of the four key features underlying the adoption profiles of the studied clusters and their classification into distinct adoption pathways.
ATI: The highly skewed ATI distribution (Sk = 1.97) with a long right tail and high kurtosis (K = 8.66) indicates that while most clusters exhibit relatively low or near mean ATI values, a subset undergoes significantly higher adoption intensities over time. These characteristics highlight the pronounced variability in adoption dynamics and suggest substantial potential for further adoption before reaching saturation.

Entry Time: The distribution of entry times follows a mildly asymmetrical pattern, with a mean adoption onset in May 2015 (SD = 2.52 years). The low kurtosis (K = 0.18) indicates a relatively uniform spread of entry times across the adoption timeline. However, the positive skewness (Sk = 1.09) suggests that a notable number of clusters were adopted later than the regional mean.

Latest Trajectory: The majority of clusters (57.6\%) exhibit a stable trajectory, meaning their curves did not cross the regional mean curve throughout the timeline. The remaining clusters are evenly split between downhill (21.6\%) and uphill (20.8\%) trajectories, indicating a high prevalence of shifts over time and substantial heterogeneity in adoption dynamics across the investigated clusters.

Latest Adoption Intensity (LAI): The LAI distribution resembles that of ATI but exhibits an even more pronounced right skew (Sk = 2.56) and heavier tails (K = 13.13) than observed in ATI. This suggests that while most clusters maintain moderate adoption intensities at the end of the studied timeline, a minority exhibit extreme values that deviate more significantly from the regional mean at that specific point in time. The high standard deviation (SD = 89.35) further reinforces the considerable variability across clusters and supports the observation of significant disparities, suggesting that the region may not yet have reached its full adoption potential.

\subsection{Adoption paths}
\begin{figure}[]  
   \setlength{\fboxsep}{0.5pt} 
    \setlength{\fboxrule}{0.5pt} 
    \fbox{\includegraphics[width=\columnwidth]{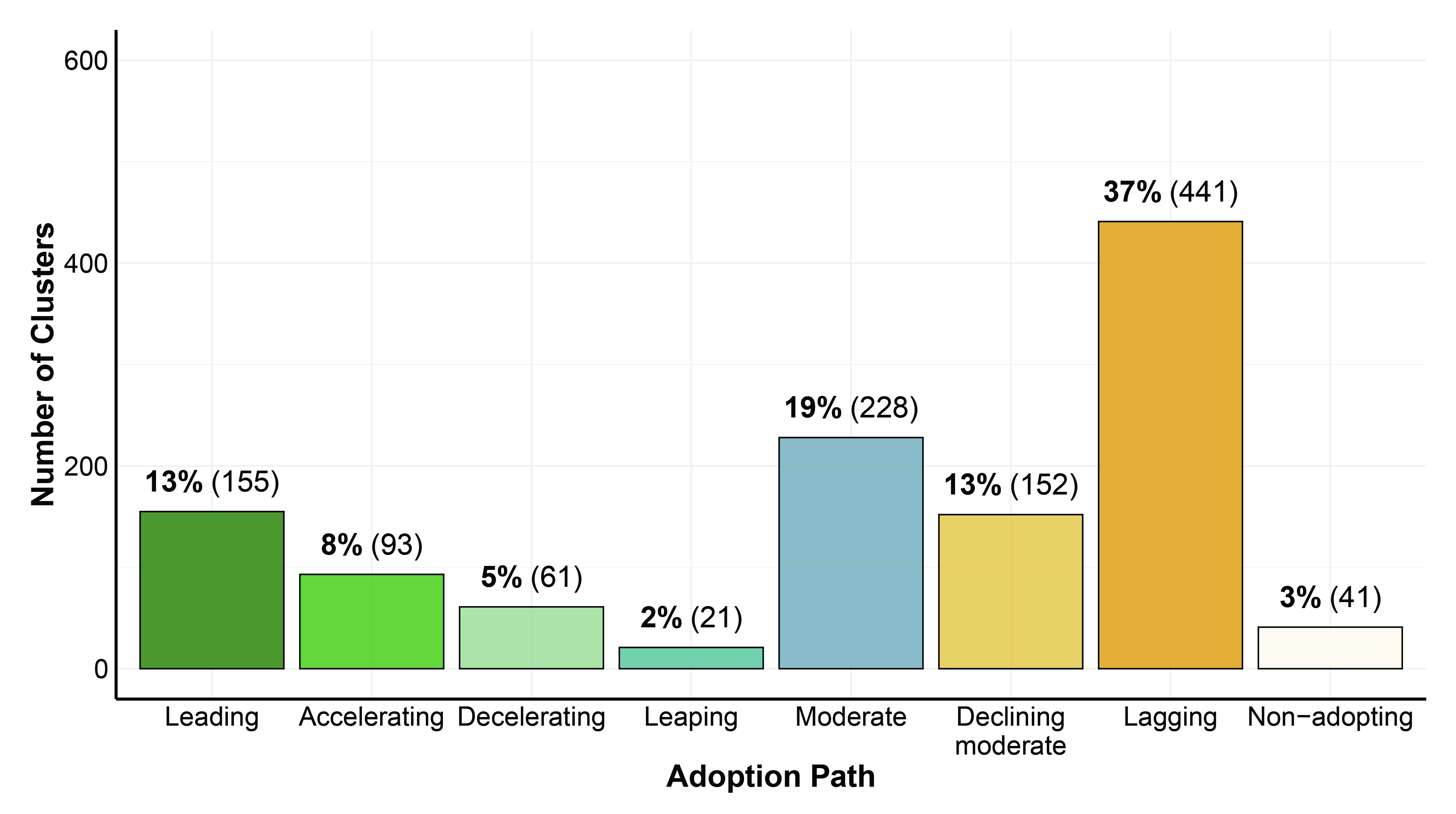}}
    \caption{\textbf{Distribution of clusters by adoption path:} The predominance of the Lagging path indicates that many clusters remain at relatively low adoption levels, suggesting that the region may not yet have reached its potential for broader diffusion. The Moderate path represents the second-largest group. The Decelerating path accounts for 5\% of the clusters, while the declining moderate path is 2.6 times larger (13\%). Namely, stagnation or a decline in adoption rates is more frequent among clusters with adoption intensities close to the mean. The Leading and Accelerating paths together account for 21\% of clusters, highlighting a notable presence of early and proactive adopters. The Leaping path is infrequent, suggesting that late but intense adoption is uncommon among the studied population.}   
    \label{paths_freq}
\end{figure}

\begin{figure}[]
    \flushleft
    \setlength{\fboxsep}{0.5pt} 
    \setlength{\fboxrule}{0.5pt} 
     \fbox{\includegraphics[width=\columnwidth]{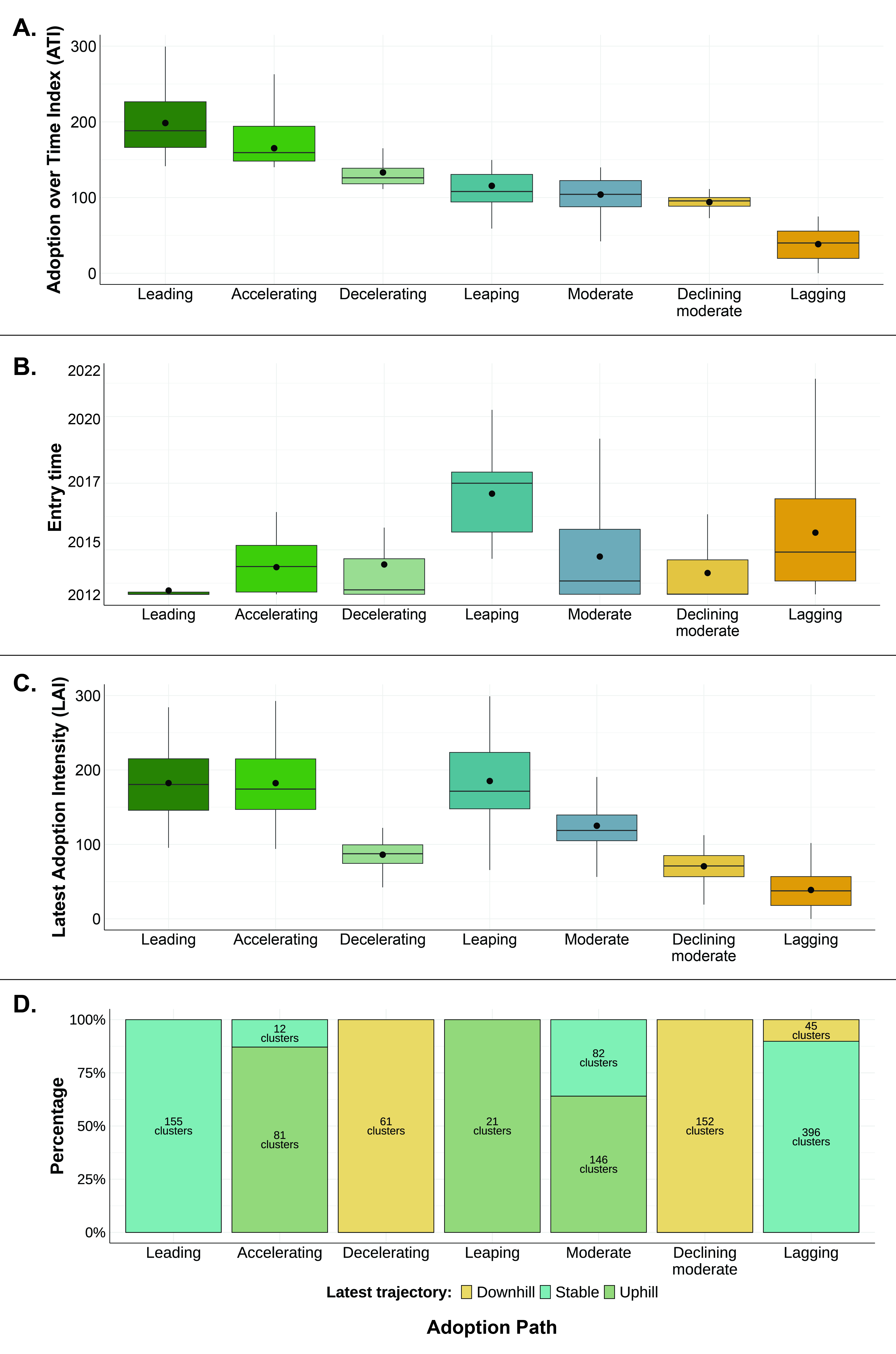}}
    \caption{\textbf{Adoption paths according to key features.} A-C: Differences in adoption paths across continuous variable scales show a clear downward trend in ATI, with significant differences across most paths. In LAI, this trend is disrupted by the Leaping path with high levels and low levels of Decelerating. LAI differences emerge between declining moderate and moderate paths. Leading and Accelerating paths show a clear mean ATI difference despite both being typologies under the same ordinal ATI category. This gap disappears in LAI. Conversely, while Decelerating and Accelerating paths show no mean ATI difference, a significant gap emerges in LAI. These findings reinforce the dynamic trends and shifts in adoption over time. See in Fig. \ref{anova} p-values from ANOVA's Tukey post-hoc test, comparing mean differences between adoption path pairs for A-C features. D: Mixed trends appear in the Moderate path, where 64\% of clusters have an uphill trajectory, indicating that the adoption process is still ongoing and many clusters have not yet reached saturation. By contrast, 10\% of the Lagging clusters exhibit a downhill trajectory, suggesting further decline even among the weakest adopters. 
}   
    \label{paths}
\end{figure}

Fig. \ref{paths_freq} and \ref{paths} illustrate the distribution of distinct adoption paths and, respectively, the characteristic values of key adoption features for each path. As seen in Fig. \ref{paths_freq}, the investigated clusters exhibit significant differences in their adoption behavior, with a notable dominance of lagging path, highlighting a considerable gap between current adoption levels and full saturation. The moderate and declining moderate paths stand out, together comprising approximately one-third of the clusters; 13\% of them exhibit stagnation or decline toward the end of the examined timeline. Among the front-runners, 13\% of the clusters display a leading behavior, namely, a consistent and sustained process of high adoption values, and 8\% have undergone an acceleration process early in the timeline. 61 clusters (5\%) decelerating over time, meaning they initially exhibit high adoption intensities but fail to maintain them, showing stagnation or decline toward the end of the timeline. Together with the Declining Moderate path, the retreating groups account for 18\% of the clusters. Three percent of the clusters do not participate in adoption processes at all. Particularly low rates are observed in the leaping (2\%), indicating that abrupt upward shifts in adoption processes are not common among the studied population.

\begin{figure}[]
    \flushleft
    \setlength{\fboxsep}{0.5pt} 
    \setlength{\fboxrule}{0.5pt} 
     \fbox{\includegraphics[width=\columnwidth]{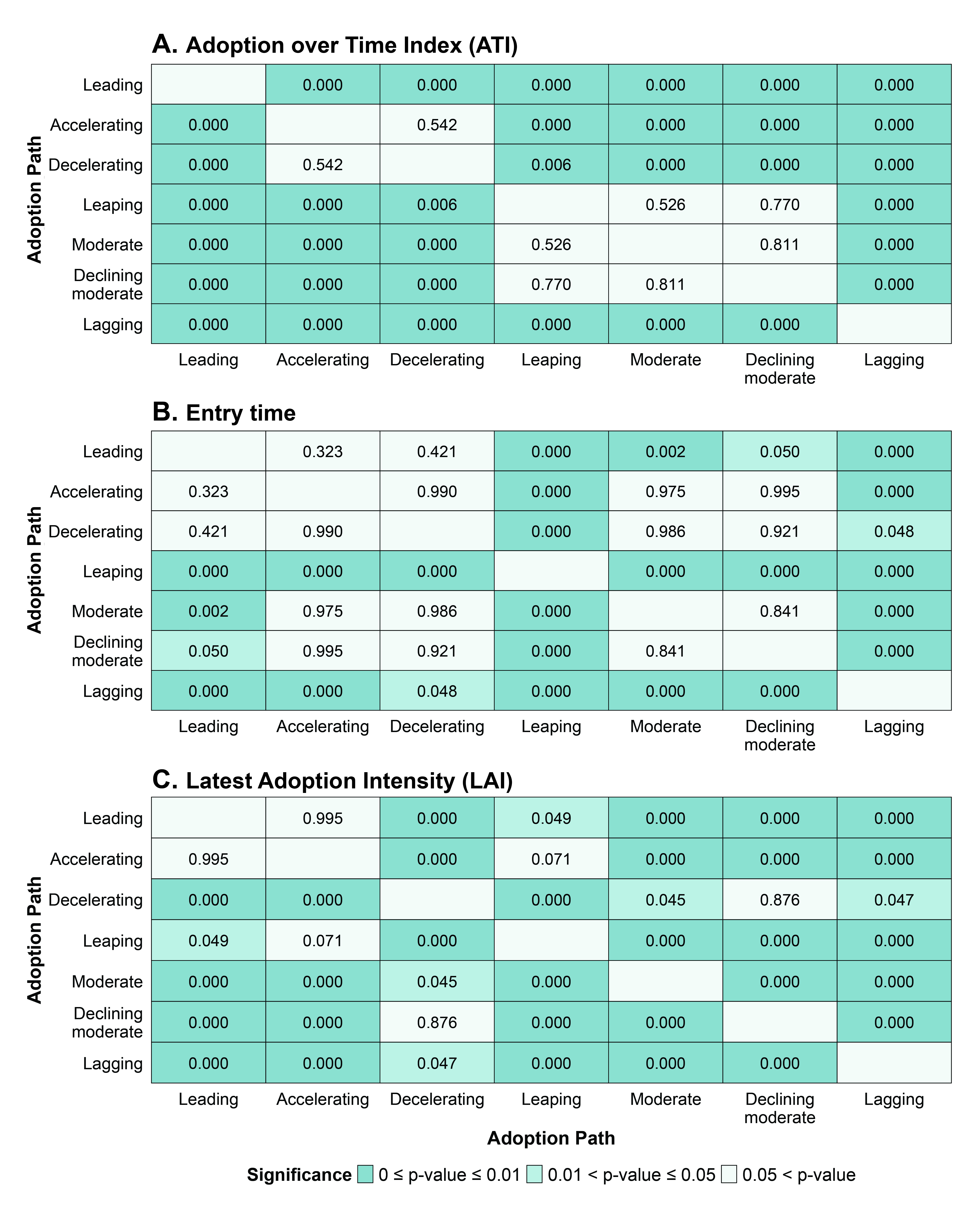}}
    \caption{\textbf{Tukey posthoc p-values} from ANOVA tests for mean differences between each pair of adoption paths by variable.}    
    \label{anova}
\end{figure}

The criterion-based typology of adoption paths relies on a broad categorization of key features. Therefore, analyzing the quantitative differences of each path along the continuous scale of these features (Fig. \ref{paths}) provides additional insights for refining the path's behavioral distinctions. Fig. \ref{anova} provides p-values of Tukey post-hoc following ANOVA tests for path differences, by feature. 

Significant differences between ATI values, across adoption paths, were observed (Figs. \ref{paths}A, \ref{anova}) even in paths categorized under the same ordinal typology value (high ATI for leading, accelerating, and decelerating; low ATI for leaping and lagging). The clear downward trend in ATI values across paths is offset when examining LAI (C) continuous values across paths, which indicate that adoption behavior evolves over time. Specifically, the leaping path exhibits high values despite having medium-to-low ATI values, whereas the decelerating path shows an opposite trend. At this point in time, differences also emerge between the declining moderate and moderate paths, which are not evident in ATI, while distinctions between leading and accelerating paths, which were prominent in ATI, disappear entirely. An inverse process is observed between accelerating and decelerating paths, where their ATI values are hardly distinguishable, but their LAI differences are substantial.

Differences between adoption paths are also evident in entry timing. The three high-adoption paths (leading, accelerating, and decelerating) exhibit variation in both mean and range of entry time. Interestingly, both retreating paths - declining moderate and decelerating - show a narrower entry time range than those of their corresponding counterparts (moderate and accelerating, respectively), suggesting a possible link between adoption degradation and early saturation at relatively low adoption intensities.

The latest trajectory of paths is presented in Fig. \ref{paths}D. While this ordinal feature serves as the basis for classifying several adoption paths, a noteworthy insight arises from the mixed trends observed in moderate and lagging paths, which are not structured around this feature. The majority of moderate clusters exhibit an upward trend (146 out of 228), indicating that progress in adoption is a defining marker among these clusters. Within the laggard clusters, while the vast majority maintain stable, approximately ten percent show a declining trend, suggesting that, at an earlier stage, these clusters managed to rise above the mean, likely for a short period as reflected in their low ATI values, yet failed to sustain this trajectory.

\begin{figure}[]
    \flushleft
    \setlength{\fboxsep}{0.5pt} 
    \setlength{\fboxrule}{0.5pt} 
     \fbox{\includegraphics[width=\columnwidth]{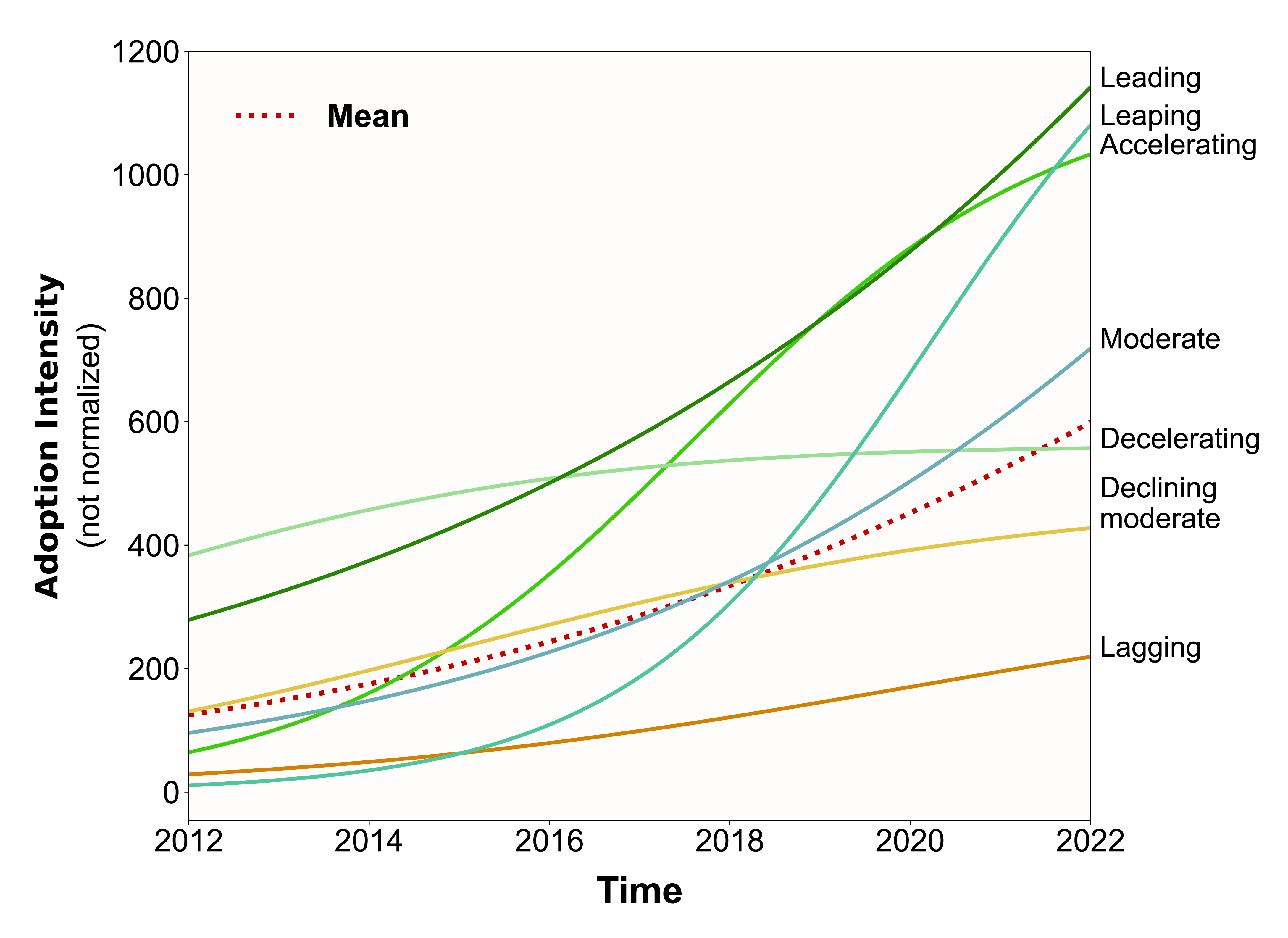}}
    \caption{\textbf{Data-driven curves of adoption paths.} Each curve represents the median adoption intensity values over time for each path, showing considerable variability that reflects diverse dynamics in adoption processes. These empirical curves strongly align with conceptual paths (Fig. \ref{theoretical_paths}) with slight deviations in the mid-range paths, which exhibit a mixed trend relative to the mean curve across the timeline.}   
    \label{curves}
\end{figure}

The data-driven curves in Fig. \ref{curves} show a strong alignment with the theoretical curves illustrated in Fig. \ref{theoretical_paths}. Thus, the implementation of an analytical framework indeed succeeds in distinguishing between adoption paths, and it reveals diverse temporal dynamics of adoption processes. A minor deviation between conceptual and data-driven curves is observed in the mid-range adoption paths. Specifically, the moderate path curve shows slightly lower adoption levels than the regional mean until 2018, after which it remains modestly above it. This pattern aligns with the finding in Fig. \ref{paths}D, which indicates that the latest trajectory of over 60\% of clusters in this group is increasing. In contrast, the declining moderate curve exhibits the opposite trend - slightly above mean adoption intensities until 2018, followed by stagnation that progressively diverges from the regional mean and ultimately falls below it. 

\subsection{Temporal transitions Between adoption paths}
Analysis of adoption path transitions over time was conducted by dividing the timeline into two equal parts, computing the key feature for each entity within each part, assigning adoption paths accordingly, and examining whether transitions occurred and, if so, what their nature and magnitude were.

\begin{figure*}[]
    \flushleft
    \setlength{\fboxsep}{0.5pt} 
    \setlength{\fboxrule}{0.5pt} 
    \fbox{\resizebox{\textwidth}{!}{\includegraphics{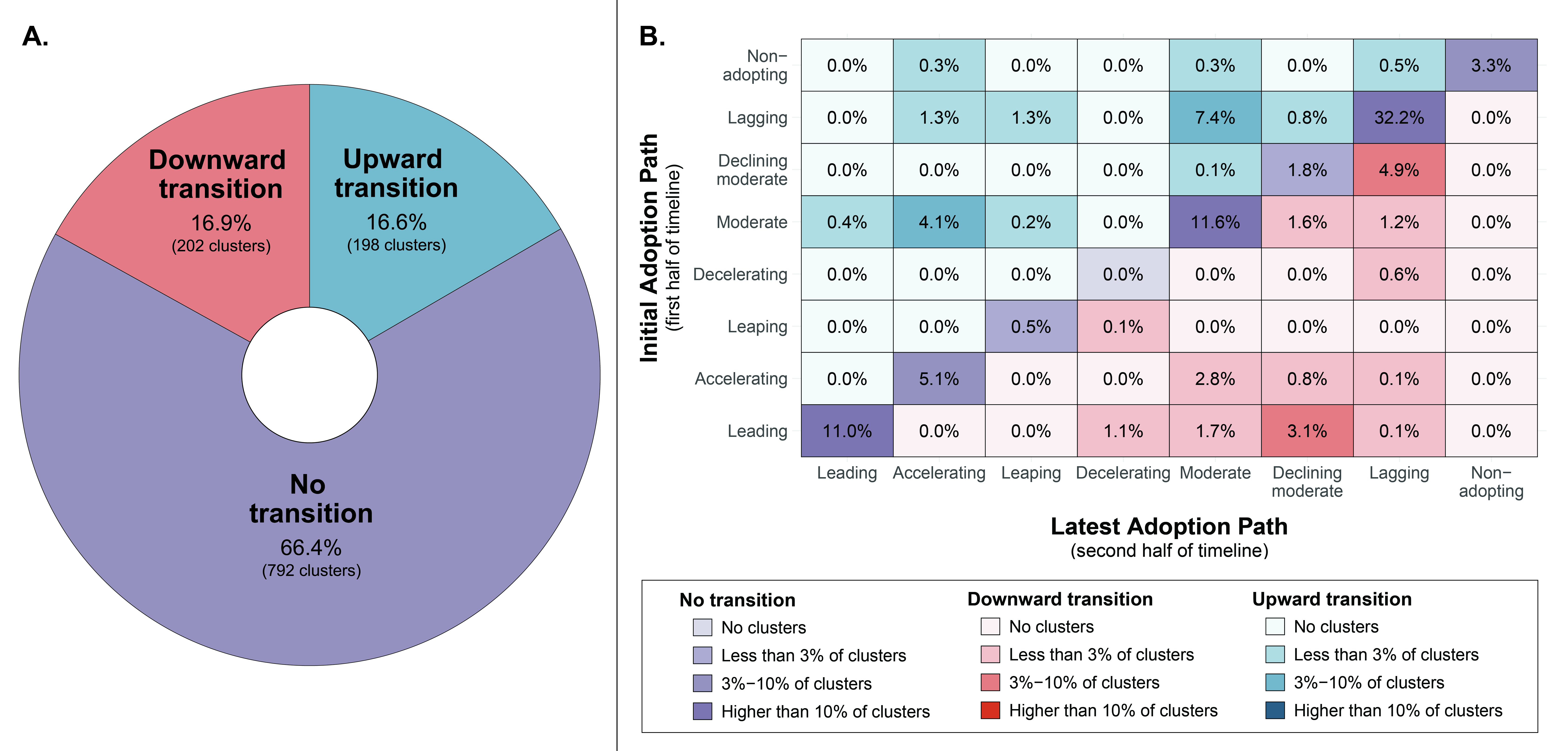}}}
    \caption{\textbf{Transitions between adoption paths from first to second timeline parts.} A: Proportion and counts of clusters transitioned downward, upward, or remained unchanged. 66.4\% clusters maintained their adoption path over time, whereas 33.6\% underwent a transition, with nearly equal proportions shifting in both directions. B: A heatmap illustrating possible transitions. Each cell represents the percentage of clusters (out of total clusters) for each combination of path pairs. The Lagging path exhibits the highest stagnation. The most frequent upward transition is in clusters shifting from the Lagging to the Moderate path, followed by clusters initially classified as Moderate and transitioned to the Accelerating path. The most frequent downward transition occurs from a Declining Moderate to a lagging path, and the most drastic downward transition occurs from the Leading path to a Declining Moderate.}   
    \label{transitions}
\end{figure*}

While most clusters (66.4\%) maintain a consistent adoption path, about one-third exhibit changes over time: One-sixth of the clusters experienced an upward transition to a higher adoption tier, while 16.9\% underwent a downward transition to a lower one (Fig. \ref{transitions}A).
Fig. \ref{transitions}B presents a heatmap of the transitions matrix, showing possible changes from each adoption path in the first part to each path in the second part. The percentages displayed along the diagonal, which represent clusters classified under the same adoption path in both parts, are the highest, indicating high temporal stability. The laggard path exhibits the greatest rigidity: 384 out of 441 clusters within this group, accounting for 32.2\% of all clusters, remained in the same category throughout the entire timeline. The moderate and leading paths follow, with approximately 12\% and 11\% of clusters, respectively, maintaining their classification over time.
Upward and downward transitions occur at lower rates across all other transition possibilities. The most frequent upward transition was observed from laggard to moderate path (7.4\% of clusters), followed by a transition from moderate to accelerating (4.1\%). A substantial jump from laggard to either accelerating or leaping occurred in 30 clusters (15 clusters, 1.3\%, in each case). The most frequent downward transition occurred from declining moderate to laggard path (58 clusters), followed by a marked shift from leading path to declining moderate, which occurred in 3.1\% of clusters (36 in total). Additionally, approximately 20 clusters transitioned from a leading to a moderate path.

\begin{figure}[]
    \flushleft
    \setlength{\fboxsep}{0.5pt} 
    \setlength{\fboxrule}{0.5pt}
     \fbox{\includegraphics[width=\columnwidth]{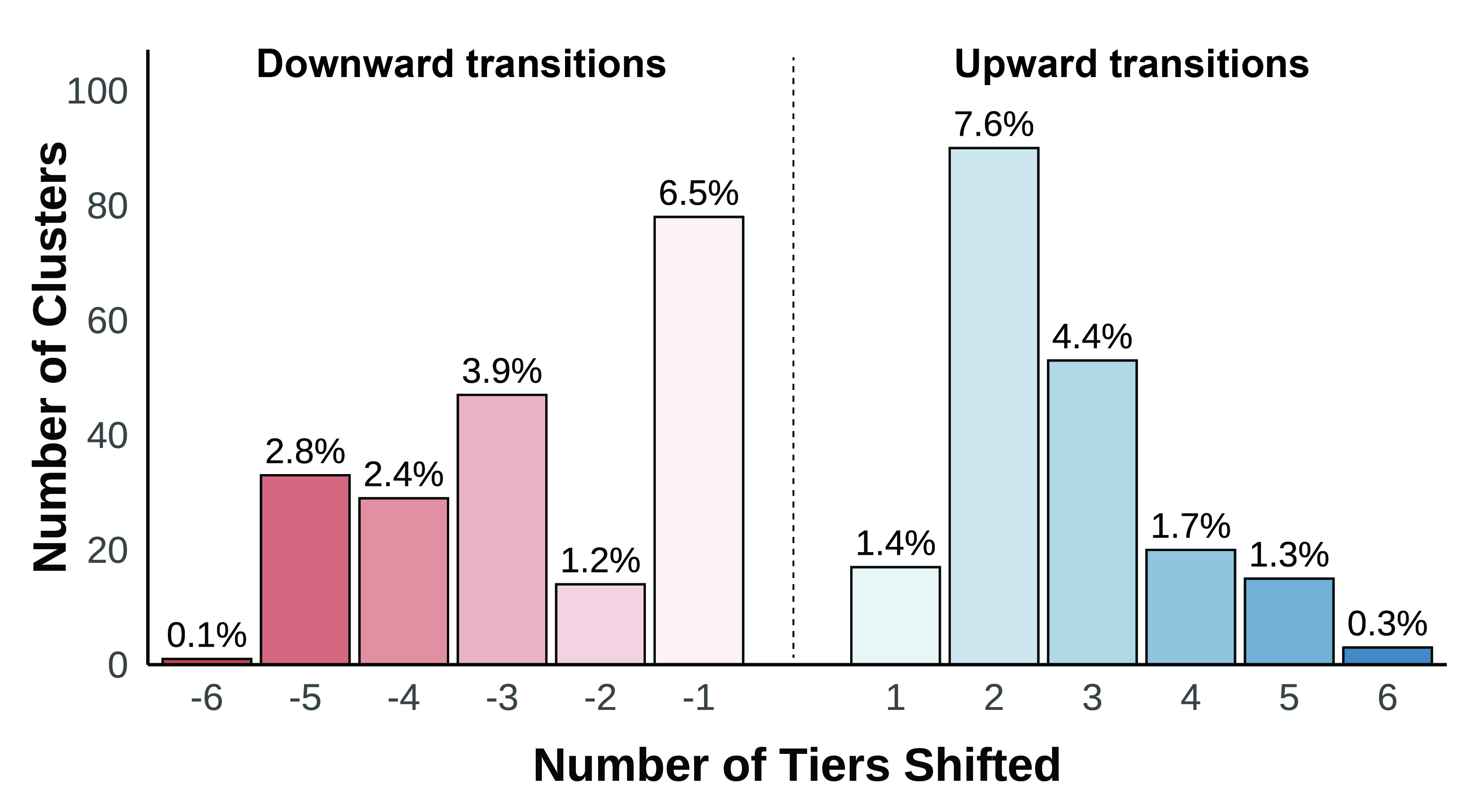}}
    \caption{\textbf{Distribution of transition magnitudes (in tiers).} Slight asymmetry is observed in transition magnitudes, with more pronounced declines compared with gradual upward shifts.}   
    \label{ranking}
\end{figure}

Fig. \ref{ranking} provides an overview of the frequency of transition magnitudes in each direction and reveals a small asymmetry between upward and downward transitions. Upward transitions are more gradual and moderate, whereas downward transitions are more abrupt. While transitions of one to two tiers occur at similar frequencies, in both directions (9\% upward and approximately 8\% downward), more drastic shifts - involving changes of four to six tiers - are more common in the downward direction, with approximately 3\% of clusters experiencing a decline of five or more tiers. This amounts to 35 clusters that exhibited more abrupt adoption slowdowns or even abandonment of efforts. In contrast, upward transitions were less sharp, with the highest frequency observed for increases of two to three tiers, totaling 143 clusters.

\section{Discussion}

\subsection{An analytical framework for adoption path typology}
The extensive literature on RE adoption highlights that transitions to RE are complex processes, shaped by several factors and differing across geographic and social contexts \citep{sovacool2016long, andersen2023faster}. Our analytical framework captures this complexity by quantifying the continuous de facto intensity and temporal dynamics of RE adoption across geographic entities within a given region, thereby enabling the identification of each entity’s distinct adoption trajectory over time. 

Unlike previous studies, this framework challenges the often questionable assumption that RE adoption is a unidirectional and accelerating process \citep{ayoub2024, geels2023socio}. This assumption is inherently embedded in many studies that rely solely on cumulative RE adoption growth functions, thereby overlooking potential phases of deceleration, stagnation, or even regression over time. For example, \citet{Morcillo2022} simulate PV adoption as a process of uninterrupted growth without considering scenarios of decline or withdrawal. Similarly, \citet{soderholm2007wind} assume a monotonic increase in installed capacity of wind power over time, overlooking possible drops in demand. Studies that do address RE adoption dynamics typically do so through long-term, large-scale, ex-ante projections, embedding adoption trends only as hypothetical scenarios rather than as empirically observed patterns. This approach is illustrated, for example, in \citet{dianat2022combining}, where the authors project RE transitions in the Middle East and North Africa up to 2060, and in \citet{cherp2021national}, which presents projections of solar and wind technology adoption at national and global scales. A contrasting example is offered by \citet{peralta2022spatio}, who uses high-resolution spatiotemporal data of the UK to train artificial neural networks, which are based on past dynamics learning to generate short-term PV forecasts. The need for time-sensitive, empirically grounded analysis that can capture diverse adoption trajectories is increasingly acknowledged. For example, the review paper of \citet{andersen2023faster} explicitly calls for approaches that allow quantifying acceleration patterns, reversals, and feedback loops over time as a prerequisite for understanding and guiding RE transitions. 

In addition to pinpointing various dynamics, our framework facilitates a regional analysis by relying on normalized quantification of adoption features against the regional average trend, which serves as a benchmark. This approach is grounded in the regional science economic perspective, which emphasizes the importance of analyzing innovation diffusion and adoption within their territorial context. This can reveal intra-regional inequalities and region-specific capacities and constraints. The works of \citet{asheim2005knowledge}, \citet{cooke2024introduction}, \citet{hoicka2025insights}, and \citet{hassink2019towards} underscore the regional system as a necessary framework for understanding innovation adoption and diffusion. These studies emphasize the importance of grounding analysis in the local context and use it as a comparative baseline to reveal dependencies and patterns. Aligned with this perspective, our approach adopts a relative assessment of entities’ adoption patterns within their respective region while referring to it as a whole - a strategy that enables a more granular perspective of local behavioral patterns.

A fundamental analytical component of the framework is the new index, ATI. It enables continuous quantification of adoption intensities over time and completes three additional milestone features along the process. ATI leverages well-established cumulative functions of innovation's diffusion and introduces two novel aspects: (i) It applies an integral-based computation, which, despite its widespread usage for dynamics measurement (e.g.,\citep{jentsch2019theory,dowd2009socio, yao2024quantifying}), was not applied yet in RE adoption studies; and (ii) It incorporates a feedback mechanism to account for temporal trend shifts throughout the process. The feedback mechanism considers both trend shifts and their duration. Consequently, ATI improves the identification of different adoption paths over time. The application of ATI to a large real-world dataset demonstrates its value as a discriminative metric, laying the foundation for a deeper understanding of the adoption dynamics of different entities. ATI reduces variance by drawing values closer to the center of the distribution, while simultaneously increasing dispersion at the extremes (Table \ref{comparison}). This enhances the differentiation between mid-range and boundary cases and enables the distinction between divergent trajectories at the extremes. For example, it enables distinguishing between an entity that enters the process at the adoption onset and maintains a high rate of adoption intensities throughout, and an entity that joins the process at a later stage, gains momentum, and reaches similar cumulative adoption intensities. By that, it positions the former closer to the leading path due to its pioneering and persistent nature. Conversely, ATI allows distinguishing between entities with low cumulative intensities, provided that one surpasses the regional mean trend at some point along the process. This dynamic increases the entity's ATI score and sets it apart from an entity that consistently remains below the regional mean. In doing so, ATI's feedback mechanism captures a fundamental difference between adoption dynamics, thereby enabling a more accurate typology than currently offered in the literature. 

\subsection{Typology of adoption paths}
Our analytical framework yielded a refined typology of eight RE adoption trajectories, designated \textit{adoption paths}. By contrast, prior frameworks have established only fundamental adoption paths, including leaders, laggards, and occasionally one or two intermediate classifications \citep{wang2022, RUOKAMO2023103183}. Within the framework innovations, three of the eight adoption paths receive particular emphasis: the \textit{decelerating}, \textit{declining moderate}, and \textit{leaping} paths. The two downward-trending trajectories constitute novel contributions of our research, while the leaping path, although widely recognized in RE literature \citep{yap2022towards, lee2020economics}, was not integrated into comprehensive adoption typologies thus far. This expansion stems from our incorporation of temporal adoption dynamics, which previous typologies typically neglected. Moreover, typologies in previous studies were mostly examined at a national
scale \citep{yap2022towards, levin2016can} and not at a regional scale as we did. In addition to these three paths in focus, our framework also refined the distinction between the two front-runner groups - \textit{leading} and \textit{accelerating}, and the two trailing groups - \textit{lagging} and \textit{non-adopting}. As a result, these eight paths enable to track and understand the essential dynamics that unfold during the adoption process.

A regional classification of geographic entities, by their adoption paths, offers insights into the dynamics of RE technology uptake by local communities and enables policy interventions tailored to the specific characteristics of each entity. The efficacy of aligning policy interventions with the specific needs of adoption groups is well-documented in the literature, e.g., \citet{vasseur2013comparative} in Japan and the Netherlands, and \citet{jacobsson2006politics} in Germany. Both studies highlight the role of context-aware and locally responsive policies for successful RE adoption. Special attention should be given to identifying groups exhibiting decline, or deceleration, in adoption over time. In these cases, external intervention is required to prevent further deterioration and entrenchment while requiring small investments compared with other groups. In these declining/decelerating entities, major adoption barriers were crossed, and observed stagnation/regression stems seemingly from local disruptions or solvable challenges. A decline in adoption rates, which may be caused by low-quality components with faster wear \citep{chowdhury2016off}, deteriorating infrastructure, and inadequate maintenance services \citet{chen2021linking}, can be mitigated through targeted local rehabilitation efforts and stricter technical standards to ensure long-term sustainability. Additionally, training service and maintenance personnel was identified as an effective strategy to reinforce adoption processes that have already taken hold, particularly in developing regions, as was demonstrated in Vietnam \citep{do2020underlying},  Colombia \citep{rodriguez2018photovoltaic}, and South Brazil \citep{garlet2019paths}.
The distinction we achieved between adoption paths enabled the identification of front-running entities that may serve as catalysts for RE adoption by activating well-documented mechanisms of imitation, and social/spatial contagion \citep{benabou2006incentives}. Such mechanisms are considered key drivers in innovation diffusion processes \citep{rogers1995lessons}, and their relevance to RE adoption was demonstrated in numerous studies, across households and neighborhoods \citep{mundaca2020drives}, farms and companies \citep{baranzini2017drives}, and even between similar localities \citep{GRAZIANO201975}. However, identifying all front-runners as “leaders” provides only a partial picture and fails to encompass the full range of early adoption trajectories. As defined in Rogers’ seminal work \citep{rogers1962} and demonstrated in subsequent studies (e.g., \citep{palm2020early}), leaders tend to be risk-takers, open to technological novelty, possess above-average technical competence, and are often driven by ideological motivations and intellectual curiosity.
These attributes may not resonate with the larger conservative adoption groups, which are typically more cautious and risk-averse \citep{kahneman2013prospect}.
Moreover, conservatives tend to be more responsive to proven benefits, particularly economic returns, which, according to the comprehensive review by \citet{shakeel2023solar}, were strongly associated with PV adoption in 72\% of the studies on adoption motivations. Hence, distinguishing between leaders and accelerators becomes particularly important. The accelerating path is characterized by a slightly later entry into the adoption process. This typically occurs in parallel with technology consolidation, increased economic feasibility, and improved access to maintenance and service infrastructures. Consequently, this group may serve as a more relatable and influential reference point for mainstream entities. Identifying the right entities with the potential to influence others and serve as emulation models can enhance the efficacy of policy measures. For example, involving local leadership from accelerating entities in outreach initiatives was found by \citet{shi2022village} to be effective in rural China. Similarly, transforming these entities into knowledge and training hubs can amplify their impact, as suggested by \citet{curtius2018shotgun} and \citet{lenton2022operationalising}. Along the same lines, identifying leaping entities, namely those exhibiting abrupt surges in adoption, can help leverage their success stories as sources of inspiration for less active entities \citep{o2022renewable}.

Distinguishing between laggards, who adopt slowly and hesitantly, and non-adopters, who opt out of the process altogether, is also essential but missing from existing works. These two groups require different engagement strategies, as observed by \citet{o2022renewable}. Laggards may be more responsive to soft bottom-up interventions. These may include enhancing community involvement, strengthening ties with high-adoption entities, supporting local cooperatives formation, which is found effective for emulation \citep{noll2014solar}, or increasing involvement of women and younger generations in the decision-making processes surrounding adoption \citep{simpson2021adoption,khan2019determinants}. Increasing adoption in non-participating entities, however, may require more assertive top-down measures, as highlighted by \citet{chen2021linking}, demonstrating the importance of these approaches in Wuhan, China. Such means can include micro-financing schemes \citep{khan2019determinants}, pay-as-you-go models \citep{montoya2022simulating}, and other economic instruments and awareness-raising measures to promote adoption \citep{aarakit2022role}.

\subsection{The off-grid Bedouin communities}
Our analytical framework was applied as a case study to the off-grid Bedouin communities of the Negev desert, Israel. It is a socioeconomically marginalized population \citep{yahel2021tribalism} residing in a vast region, in small and dispersed rural clusters of households, and facing persistent and severe energy poverty \citep{teschner2020extreme}.
Identifying adoption paths of the geographical entities (clusters) of this society through our typology can be a key step towards assessing both the extent and disparities of PV adoption in the region. This allows for revealing underlying dynamics and estimating future growth potential. It also enables identification of vulnerable groups, with limited access to energy sources, entities with declining adoption requiring targeted intervention, and front-runners who can serve as role models to promote broader regional adoption. These insights can facilitate measures to enhance equitable energy access, mitigate energy poverty \citep{rao2022assessment}, and strengthen resilience \citep{shapira2021energy}.

Since we focus on adoption dynamics, identifying an adequate time frame is crucial. We identified 2012 as the starting point, because the reduction in PV prices around this year initiated the feasibility of solar solutions \citep{IRENA2024a}, even for basic needs, such as lighting and charging \citep{WorldBank2024}. Previously, high prices had rendered such technologies unattainable due to low economic power. Consequently, this 10-year timeline enables an examination of the adoption process from its inception, when the technology had just become attainable, but early adopters still faced limited availability of distributors and service providers.

Typology analyses revealed large regional disparities, with trailing groups comprising 40\% of the clusters, surpassing the share of mid-range groups. While this outcome diverges from expectations of greater concentration around the regional mean curve, after a decade of PV deployment, it is nonetheless consistent with some other studies of off-grid rural areas. For example, evidence from Uganda \citep{aarakit2021adoption} and India \citep{akter2021off} similarly reveals substantial regional disparities in PV adoption and a predominance of lagging groups.

Moreover, large disparities in adoption intensities between front-runners and mid-range groups, observed both cumulatively and at the most recent data point, indicate that these gaps still persist and have not diminished over time. This suggests that the regional adoption potential has not been fully realized and that the process did not reach maturity.

We found a notable representation of retreating groups (18\%), indicating that stagnation/deterioration is not merely possible after adoption was established, but is rather common. Analyzing path transitions across the two parts of the timeline reinforces this observation, showing that in 17\% of the clusters, relative adoption rates and intensities shifted downward towards the end of the studied period. As was observed in other off-grid communities (e.g., in China \citep{chen2021linking}, Bangladesh \citep{chowdhury2016off}, and Côte d'Ivoire \citep{diallo2020effects}), such declines can stem from various causes. These include failures of system components over time, infrastructure degradation, and insufficient maintenance and service systems to provide long-term support. These findings underscore the need for proactive measures to halt regression and prevent its entrenchment. 

Unlike national-scale studies \citep{hastie2024independence}, suggesting that leapfrogging is a viable trajectory for marginal entities, a significant leap in late adopters is rather rare in the Bedouin case, accounting for only 2\% of the clusters. This indicates that, despite the favorable conditions resulting from established knowledge and proven feasibility, developed service availability, and a continuous decline in prices, late adopters tend to adopt at a slower pace, and only a few capitalize on these advantages to achieve intensive deployment that positions them high above the regional average. This finding points to the need to reconsider optimistic projections of rapid RE diffusion and highlights the importance of more targeted incentive measures to accelerate adoption among entities that have initiated adoption but continue to progress at a slow pace.

Due to limited access to reliable data on off-grid populations, previous studies have typically relied on sample-based self-reported surveys to estimate solar energy use or willingness to adopt (e.g., \citep{tetteh2022determinants, aarakit2021adoption, akter2021off}, for Ghana, India, and Uganda, respectively). While sample-based surveys provide valuable insights, they suffer from inherent limitations in generalizability and are susceptible to various forms of bias, including self-selection and measurement errors. By offering a comprehensive temporal mapping of PV presence, this study fills a major gap in the literature and overcomes limitations of prior research. This data-driven approach, based on complete spatial coverage and actual adoption outcomes, enables precise quantitative measurement of the extent and evolution of PV uptake - an effort not previously undertaken in off-grid regions.

\subsection{Limitations and Future Directions}
While offering valuable analytical contributions, several limitations and avenues for future research should be noted.

First, while the proposed typology effectively identifies actual data-driven adoption paths, namely, capturing what is occurring in practice, it does not model the causal mechanisms behind these trends. Further research into the relationships between the unique characteristics of entities exhibiting different adoption dynamics, and the determinants shaping them, is needed to develop a deeper understanding of the observed variability.

Second, our framework relies on the availability of sufficiently high-resolution spatial and temporal data. The ability to generate meaningful insights depends on access to such resources, which may be limited in regions lacking continuous or detailed records, such as registries of RE system deployments. However, with the increasing use of high-resolution remotely sensed data and deep learning techniques, these limitations are expected to be reduced.

Third, although the framework is applicable to varying time spans and does not require coverage of the entire adoption life-cycle, (thereby enabling even early-stage identification of potential deceleration or decline), it nonetheless requires careful consideration of the temporal scope when interpreting the dynamics it reveals, since some aspects, including trend stability over time, may only become apparent through longer observation periods.

The proposed framework offers broad applicability across regions, spatial scales, and stages of the adoption process, providing a fertile basis for future comparative research. Such studies could investigate differences in adoption dynamics between developed and developing regions, markets with varying economic structures, levels of electricity access, degrees of social connectedness, and differing environmental perceptions and values. A comparative approach can yield valuable insights into the contextual factors shaping RE adoption diffusion.

Moreover, the suggested typology provides a solid foundation for exploring spatial patterns in adoption dynamics and offers opportunities for future research to investigate autocorrelation between different paths, including spillover effects and the potential for diffusion across neighboring entities and its range.

\section{Conclusions}
This study proposes a new analytical framework for quantifying RE adoption processes and tracing distinct adoption dynamics over time. It supports a systematic classification of every geographic entity in a given region based on its measured adoption path. 
Such a typology represents a critical step toward unraveling adoption patterns by disentangling the underlying behavioral dynamics of adopters. It also provides a foundation for policymakers to devise targeted strategies that reflect specific characteristics and needs of each group, ultimately promoting the deployment of RE technologies. This approach is crucial for aligning with global climate goals and improving human well-being, especially in off-grid marginalized areas that often suffer from severe energy poverty.

Four key conclusions emerge from the framework application:
\begin{itemize} [leftmargin=1em, itemsep=0pt]
    \item Four key features were identified as pivotal in differentiating adoption dynamics over time. The key features collectively represent essential components of adoption processes: Entry timing, reflecting innovation affinity; cumulative adoption through time, including trend shifts along the way, indicating both the intensity of the process and its stability/volatility; the most recent trajectory direction, signaling likely future progression; and the deployment intensity at the end of the timeline, depicting the most current state of RE adoption. Integration of these features effectively captured adoption processes' complexity and was found to be indicative in discerning between data-driven adoption paths.
    
    \item The new index, ATI, was developed to quantify cumulative adoption intensity over time while accounting for shifts in adoption trends throughout the process. ATI provides flexibility in capturing a wide range of adoption trajectories that existing metrics have not adequately addressed. It enhances distinguishability between fundamental adoption dynamics within both front-runner and trailing groups, as well as identifying retreating groups.
    
    \item A comprehensive typology was identified, encompassing eight distinct RE technology adoption paths over time, three of which are absent from existing classifications: \textit{decelerating}, \textit{declining moderate}, and \textit{leaping} paths. A data-driven implementation of this typology revealed that retreating trends in the adoption process over time are present to a noticeable extent. Identifying entities that follow such trajectories is crucial for preventing backsliding and addressing stagnation in the diffusion process. Moreover, the typology distinguishes between fundamental adoption dynamics, thereby improving the efficacy of measures to foster adoption by tailoring a strategy to each group's characteristics.
    
    \item The off-grid Bedouin population in southern Israel displays pronounced disparity in PV adoption intensities, with a substantial presence of lagging and regressing groups. Despite a decade of access to solar solutions, adoption intensities remain below the region’s potential, clearly indicating that this period is insufficient to achieve widespread adoption, let alone full saturation. These findings underscore the need for vigorous interventions to accelerate PV adoption and alleviate these communities’ energy poverty.    
\end{itemize}

\section*{Acknowledgment}
The authors thank the annotators who participated in the project. This work was supported by Israel
Science Foundation (ISF) under Grant 299/23.

\section*{Data Availability Statement}
All data and Python code supporting the findings of this study are available from the corresponding author and will be shared upon reasonable request, without restriction.

\section*{CRediT Author Statement}
Roni Blushtein-Livnon: Conceptualization, Methodology, Software, Formal analysis, Investigation, Data curation, Writing – original draft, Visualization, Validation. Tal Svoray: Conceptualization, Methodology, Formal analysis, Writing – review and editing, Validation, Supervision, Funding acquisition. Itai Fischhendler: Conceptualization, Writing – review and editing, Supervision, Funding acquisition. Havatzelet Yahel: Conceptualization, Writing – review and editing, Supervision, Funding acquisition.Emir Galilee: Conceptualization, Writing – review and editing. Michael Dorman: Data curation, Software.

\bibliographystyle{elsarticle-harv}
\bibliography{refs}

\newpage

\bio{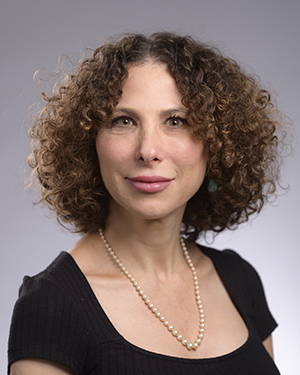}
\noindent\textbf{Roni Blushtein-Livnon} is a Ph.D. student in the Department of Environmental, Geoinformatics, and Urban Planning Sciences at Ben-Gurion University of the Negev. Her doctoral research involves, among other things, Computer Vision and focuses on deep learning methods for object detection and mapping.
\endbio
\vspace{0.8em}
\bio{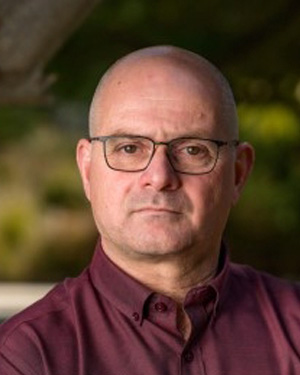}
\noindent\textbf{Tal Svoray} received a Ph.D. in Radar Remote Sensing of Mediterranean Vegetation from Bar Ilan University, Ramat-Gan, Israel, in 2001. He is currently a Professor at the Ben-Gurion University of the Negev. His main research interests include object segmentation and detection, remote sensing of soil and vegetation, environmental psychology, and geostatistics.
\endbio
\vspace{-0.4em}
\bio{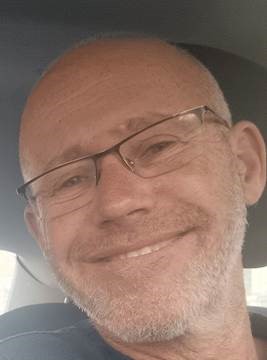}
\noindent\textbf{Itay Fischhendler} is a professor at the Hebrew University of Jerusalem. His research interests focus on environmental conflict resolution, natural resources governance, and decision-making under conditions of political and environmental uncertainties. He is a leading scholar on transboundary water and energy institutions and Middle Eastern policy. 
\endbio
\vspace{0.5em}
\bio{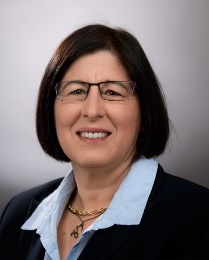}
\noindent\textbf{Havatzelet Yahel} received a Ph.D. in Historical Geography from the Hebrew University of Jerusalem, Israel, in 2016. She is currently an Associate Professor at the Ben-Gurion University of the Negev. Her main research interests include land use policies and legal and historical geography. 
\endbio
\vspace{2.2em}
\bio{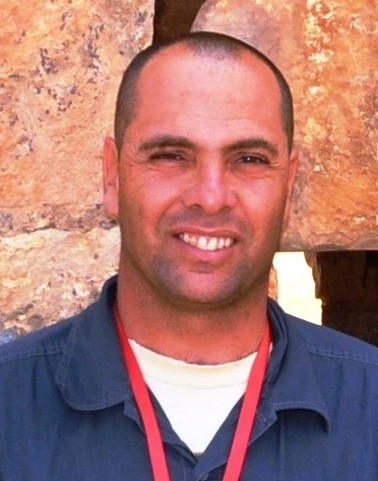}
\noindent\textbf{Emir Galilee} is an associated scholar at the Ben-Gurion Research Institute for the Study of Israel \& Zionism, Ben-Gurion University of the Negev, and a lecturer at Kaye Academic College. His areas of interest include historical, social, and cultural geography: the socio-cultural geography of minorities, nomadic and indigenous groups; Arab landscapes in historical perspective; and hiking trails and leisure landscapes.
\endbio
\vspace{-1.3em}
\bio{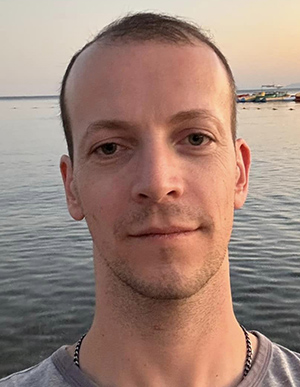}
\noindent\textbf{Michael Dorman,} Ph.D., is a programmer and lecturer at the Department of Environmental, Geoinformatics and Urban Planning Sciences, Ben-Gurion University of the Negev. He is working with researchers and students to develop computational workflows for spatial analysis, mostly through programming in Python, R, and JavaScript, as well as teaching those participants. 
\endbio

\end{document}